\documentclass{cpeauth}
\usepackage{mathptmx}
\usepackage{amsmath}
\usepackage{amssymb}
\usepackage{amstext}
\usepackage{color}
\usepackage{rotating}
\usepackage{graphicx}
\usepackage{url}
\usepackage{sidecap}
\usepackage[numbers,square]{natbib}
\usepackage{subfig}
\usepackage[breaklinks,colorlinks,linktocpage,final]{hyperref}
\hypersetup{urlcolor=blue}
\usepackage{listings}
\newcommand{\arxiv}[1]{\url{http://arxiv.org/abs/#1}}

\newcommand{\bq}{\begin{equation}}
\newcommand{\eq}{\end{equation}}
\newcommand{\bi}{\begin{itemize}}
\newcommand{\ei}{\end{itemize}}

\newcommand{\flops}{\mbox{flops}}

\newcommand{\GBS}{\mbox{GB/s}}

\newcommand{\GFLUPS}{\mbox{GFLUP/s}}
\newcommand{\GHZ}{\mbox{GHz}}
\newcommand{\W}{\mbox{W}}

\newcommand{\FLUP}{\mbox{FLUP}}
\newcommand{\FLUPs}{\mbox{FLUPs}}

\newcommand{\bytes}{\mbox{bytes}}
\newcommand{\byte}{\mbox{byte}}
\newcommand{\GB}{\mbox{GB}}

\newcommand{\eos}{~.}
\newcommand{\cma}{~,}

\newcommand{\supermuc}{Super\-MUC}
\DeclareCaptionType{copyrightbox}
\begin{document}
\runningheads{Wittmann, Hager, Zeiser, Treibig, Wellein}{Energy-optimized lattice-Boltzmann simulations}
\title{Chip-level and multi-node analysis of energy-optimized lattice-Boltzmann CFD simulations}

\author{M. Wittmann, G. Hager\corrauth, T. Zeiser, J. Treibig, and G. Wellein}
\address{Erlangen Regional Computing Center (RRZE), Martensstr. 1, 91058 Erlangen, Germany}
\corraddr{Erlangen Regional Computing Center (RRZE), Martensstr. 1, 91058 Erlangen, Germany. E-mail: \url{georg.hager@fau.de}}


\begin{abstract}

Memory-bound algorithms show complex performance and energy consumption
behavior on multicore processors.  We choose the lattice-Boltzmann
method (LBM) on an Intel Sandy Bridge cluster as a prototype 
scenario to investigate if and how single-chip performance and power
characteristics can be generalized to the highly parallel case.
First we perform an analysis of a sparse-lattice LBM implementation for
complex geometries.  Using a single-core
performance model, we predict the intra-chip saturation characteristics
and the optimal operating point in terms of energy to solution as a
function of implementation details, clock frequency, 
vectorization, and number of active cores per chip. 
We show that high
single-core performance and a correct choice of the number of active cores
per chip are the essential optimizations for lowest energy
to solution at minimal performance degradation. 
Then we extrapolate to
the MPI-parallel level and quantify the energy-saving potential of
various optimizations and execution modes, where we find
these guidelines to be even more important, especially
when communication overhead is non-negligible.
In our setup we could achieve energy savings of 35\% 
in this case, compared to a naive approach. 
We also demonstrate that a
simple non-reflective reduction of the clock speed leaves most of
the energy saving potential unused.

\end{abstract}

%
%

\maketitle

\section{Introduction}\label{sec:intro}

\footnotemark{}\footnotetext{This is the peer reviewed version of the following article: ``Wittmann, M., Hager, G., Zeiser, T., Treibig, J., and Wellein, G. (2015), Chip-level and multi-node analysis of energy-optimized lattice Boltzmann CFD simulations. Concurrency Computat.: Pract. Exper., doi: 10.1002/cpe.3489'' which has been published in final form at \url{http://dx.doi.org/10.1002/cpe.3489}. This article may be used for non-commercial purposes in accordance with Wiley Terms and Conditions for Self-Archiving.}%
For many years, high performance computing has focused solely on optimizing the
sustained performance in order to reduce the
time to solution. Owing to the increasing energy consumption and
energy costs of HPC systems, additional metrics such as
energy to solution are brought into focus. Adapting the processors'
clock frequency to the needs of an implementation is one way to influence
the energy consumption of a compute node and also to meet power capping
requirements. However, the complex topology
of modern compute nodes with typically at least two sockets and
multi-core processors as well as the different performance bottlenecks
-- such as memory bandwidth, instruction throughput, and arithmetic units
-- and their saturation behavior make an a priori prediction
difficult. Some of these bottlenecks, in particular the sustained
memory performance, often depend on the processors' clock frequency.

In this paper, we use a lattice-Boltzmann method (LBM) from
computational fluid dynamics (CFD) as a typical example for the large
class of algorithms with low computational intensity.  Despite its
seeming simplicity and ease of implementation, optimizing LBM on 
recent hardware platforms and for different application cases has been
the subject of intense research in the last ten years
\cite{schulz-2001,pan-2004,wang-2005,wellein-2006,bernaschi-2008,mattila-2008,bailey-2009,vidal-2010,zudrop-2012,wittmann-2012-cam}. Here, we conduct a thorough analysis of
performance and energy to solution on the chip and highly
parallel levels for an MPI-parallel implementation of LBM. 
We start from observations of the intra-chip saturation 
characteristics of two different implementations,
which differ in the order in which the flow data in the lattice
sites is updated (``propagation methods'' \cite{wittmann-2012-cam}).
Then we apply the 
execution--cache--memory (ECM) performance model and a simple
multi-core power model to describe the
optimal operating point in terms of performance and energy to solution 
as a function of the clock frequency and the single instruction multiple data
(SIMD) vectorization. To find out whether the knowledge thus gained 
at the chip level can be generalized to the highly parallel case,
we conduct scaling experiments on a modern
cluster system up to a point where MPI communication overhead becomes
significant.

This paper is organized as follows. The remainder of
Sect.~\ref{sec:intro} covers related work, the basics of the
lattice-Boltzmann implementations, the hardware used for testing,
and a list of contributions.
Sect.~\ref{sec:ecm} then introduces, applies, and validates the ECM model 
on the Intel Sandy Bridge
architecture. In Sect.~\ref{sec:power} we use a recently introduced
multi-core power model to identify the optimal operating points
on the chip. Finally, Sect.~\ref{sec:parallel} presents performance
data for highly parallel runs and analyzes the impact of the
different parameters (clock speed, number of cores per chip, 
SIMD vectorization, system baseline power). Sect.~\ref{sec:conclusion}
gives a summary and an outlook to future research.

\subsection{Related work}

The \textit{roof\/line} model of \textsc{Williams} et
al.~\cite{williams-2009} provides valuable insight into how much
performance can be achieved with regard to the typical limiting
factors, i.\,e.\ memory bandwidth and arithmetic throughput.  This allows for a
first assessment of how far the performance of the code at hand
deviates from the maximum sustainable performance (the ``light speed'').

Performance modeling and prediction especially in the context of LBM,
are an ongoing research topic of many groups in engineering and
computer science~\cite{lba:peters:2010,hpc:carter:2005}. Auto-tuning
was used, e.\,g., in \cite{williams-2009-lbm} for a
magnetohydrodynamics LBM. The highly optimized LBM code we use in the present
work shows sustained performance close to the roof\/line predictions,
and will be described in Sect.~\ref{sec:lbm}.

Power supply and cooling account for a significant fraction of the
total cost of ownership (TCO) of modern HPC resources. Hence, the
simplistic ``node-hour'' cost model still used in most centers
is becoming inappropriate, and there is ongoing
research towards more suitable metrics. The simplest such metric
is ``energy to solution,'' i.e., the energy required by the hardware
to solve a given problem \cite{keller:ets2012,bode:ets2013}. However,
since minimal energy to solution does not automatically imply
minimal time to solution, combined metrics such as the energy-delay
product or variants thereof have been devised 
\cite{bekas:2010,Hsu:2012:TES:2188286.2188309}. It is a matter 
of policy which metric should be the optimization target and
there is no general consensus.

Research in the direction of energy-saving hardware and software
mechanisms focuses on models and algorithms for dynamic voltage and
frequency scaling (DVFS) and dynamic concurrency throttling
(DCT)~\cite{10.1109/TPDS.2012.95}.  Frequency scaling 
changes the computational performance of a core, but it can 
also influence cache and memory bandwidth~\cite{hager:cpe13,schoene-2012}.  
Any realistic modeling
effort must take these effects into account.  The ECM model
\cite{hager:cpe13} is a refinement of the roof\/line model and
allows a more accurate prediction of the single-core performance and
the intra-chip scaling properties
of a parallel code. In \cite{hager:cpe13} we have used it to model
a specific LBM benchmark kernel.
Together with a simple power model derived for the Sandy Bridge chip 
we were able to explain the performance saturation and energy to solution
characteristics of this solver. The analysis in this paper goes much
deeper, and considers a more advanced propagation method as well as
the full-scale simulation code. It also
extends beyond the single chip to fathom the energy and performance
properties of the solver in distributed-memory parallel runs on up to
128 nodes.

Modeling the power dissipation of chips has received much attention in
recent years. A power model similar to the one used in this paper has
been introduced in \cite{choi:2012}. It concentrates, however, on
microscopic quantities such as the energy cost for transferring data
or for performing floating-point operations, and does not address
multi-core issues. It would be interesting to reconcile both models
for a more holistic view on chip power, but this is left for future
work.

\subsection{The lattice Boltzmann method}\label{sec:lbm}

Lattice-Boltzmann methods have become a popular approach in
computational fluid dynamics. However, they are also interesting for
computer scientists as the core algorithm is rather short and uses a
multi-streaming loop kernel, resulting in many concurrent memory
streams and no reuse of data in a single iteration, but is
straightforward to parallelize due to simple next-neighbor
communication.

\begin{figure}
\centering
\includegraphics*[width=0.4\textwidth]{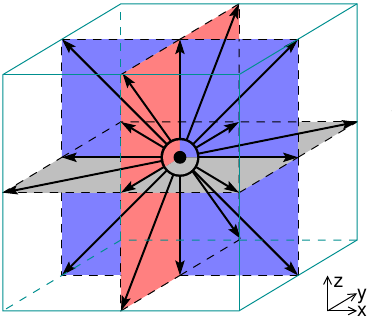}
\caption{\label{fig:d3q19}Visualization of a single lattice node with
  its $19$ stored values of the distribution functions (arrows). In
  one complete time step (``sweep''), all distribution functions at
  all lattice nodes are updated.}
\end{figure}

In the present work we employ the \emph{ILBDC} code
\cite{zeiser-2009-ppl}, which uses a D$3$Q$19$ lattice model and a
two-relaxation-time (TRT) collision model
\cite{ginzburg-2008}. D$3$Q$19$ is the most popular discretization
scheme for LBM in 3D. All calculations are performed in
double-precision floating point arithmetic.
The algorithm with the D$3$Q$19$ model can be viewed as a $19$-point stencil
in 3-D accessing only nearest neighbors but has two important differences to common 
stencil algorithms:
(i) Each lattice node consists not only of one, but of $19$ values (so called particle 
    distribution functions [PDFs]);
(ii) Each PDF accessed is only accessed again in the next time step, which prevents
    data reuse (unless complex temporal blocking schemes are employed).
Figure~\ref{fig:d3q19} shows a single lattice
site with $19$ stored  distribution functions (arrows).
The performance of a given LBM approach depends at least on the
data layout and memory access patterns, the strength of arithmetic
operations (i.\,e., how well numerical expressions are simplified and
combined to avoid unnecessary operations), and their degree of SIMD
vectorization.
A thorough overview of more propagation-step implementations and their memory
access characteristics can be found in \cite{mattila-2008, wittmann-2012-cam}.

It seems natural to store the PDFs in a 4-D array, and 
use an additional Boolean array to distinguish fluid nodes
from obstacle nodes.
This is known as the \textit{marker-and-cell} approach.
However, LBM simulations of domains with a large fraction of solid
nodes can benefit from a \textit{sparse representation} of the
domain~\cite{schulz-2001, pan-2004, wang-2005, bernaschi-2008,
  vidal-2010, zudrop-2012}, where only the fluid nodes are kept in a
1-D vector.  Indirect accesses to PDFs of neighboring nodes are then
required and accomplished through an adjacency list (IDX), which
represents the topological connections of the nodes.  ILBDC uses such
a sparse representation.

For updating one node, optimized implementations read one PDF from
each of the $18$ surrounding neighbors and the local node (streaming
step), compute updated values (collision step), and write the results
to the PDFs of the local node.  This is known as the \textit{pull}
scheme~\cite{wellein-2006}. It is implemented in the ILBDC code
together with a \textit{structure-of-arrays} (SoA) data layout where
all PDFs of a direction are stored consecutively in memory before the
next direction follows: \verb.f(i,j,k,PDF)..

To work around the data dependency problems of a combined
stream-collide step, two lattices are often used, one as the source
and one as the destination. Then, a fluid lattice-node update (\FLUP) requires
$19$ PDF loads, $19$ additional PDF loads because of
\textit{write-allocate} (also known as \emph{fetch-on-write})
transfers, $19$ PDF stores and $18$ IDX loads
of the adjacency list.
Assuming double-precision floating-point numbers (eight bytes) for PDF and
four-byte integers for IDX, the total number of bytes that must be transferred between
CPU and memory for one \FLUP\ is
$3 \times 19 \times 8 \text{~(PDF)}+ 18 \times 4 \text{~(IDX)}$\,\bytes\ $= 528$\,\bytes.
The write-allocate is triggered by the cache hardware on store misses
and loads the data at the accessed location (the complete cache line, to be exact)
from memory into the cache before it is updated.
With \textit{non-temporal store} (NT) instructions these write-allocates
are avoided and the data is directly written from the processor into
memory, bypassing the cache hierarchy.
The number of bytes required for one \FLUP\ decreases then to
$2 \times 19 \times 8 \text{~(PDF)}+ 18 \times 4 \text{~(IDX)}$\,\bytes\
$= 376$\,\bytes.
Current standard processors have difficulties with $19$ concurrent write
streams, in particular if they consist of NT stores.  As a
remedy, blocking can be applied so that a node's PDFs are read in chunks
and updated values are stored in a small temporary buffer, which
should be small enough to fit in the L1 cache.  From this buffer,
two directions of the updated PDFs at a time are written 
to the destination lattice.  We call this
implementation \textit{pull-split T} or \textit{pull-split NT},
depending on whether normal (``temporal'') stores or 
non-temporal stores are used.

\begin{figure}[tbp]
  \begin{center}
    \subfloat[Even time step: read PDFs.]{
      \label{fig:aa:l1}
      \includegraphics[width=0.20\textwidth, clip=true]{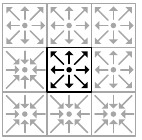}
    }
    \hfill
    \subfloat[Even time step: write PDFs.]{
      \label{fig:aa:l2}
      \includegraphics[width=0.20\textwidth, clip=true]{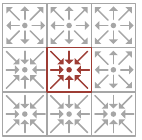}
    } \,
    \hfill
    \subfloat[Odd time step: read PDFs.]{
      \label{fig:aa:r1}
      \includegraphics[width=0.20\textwidth, clip=true]{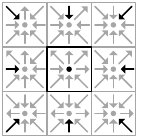}
    }
    \hfill
    \subfloat[Odd time step: write PDFs.]{
      \label{fig:aa:r2}
      \includegraphics[width=0.20\textwidth, clip=true]{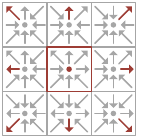}
    }
    \caption{Visualization of the AA pattern (or propagation model),
      using the D$2$Q$9$ model for the sake of clarity. In the even
      time step of the AA pattern, only PDFs of the current node are
      read~(a), collided, and written back to the same
      location~(b) but into opposite directions.  As a
      consequence, the current node PDFs are located at neighboring
      nodes during the following odd time step. These are then loaded
      from the opposing direction~(c), collided, and
      written back~(d).  }
    \label{fig:aa}
  \end{center}
\end{figure}

\textsc{Bailey's} \textit{AA pattern} \cite{bailey-2009} for the PDF
access allows using one single lattice only (instead of separate
source and destination grids) while maintaining the possibility to
update all cells in any order and in parallel.  The iterations over
the lattice are divided into even and odd time steps.  During an even
time step (Fig.~\ref{fig:aa:l1} and~\ref{fig:aa:l2}) only PDFs of the
current node are accessed in each lattice site update.  In the
following odd time step (Fig.~\ref{fig:aa:r1} and~\ref{fig:aa:r2})
only PDFs of the neighboring nodes are accessed, which requires
indirect addressing in our case of the sparse representation.
This update scheme only performs stores to locations in memory which have
previously been read.  No write-allocate is necessary, since the data
to be updated already resides in the cache.
We use an optimized version where the even time step is completely
SIMD-vectorizable, which can easily be accomplished as all PDFs
are accessed consecutively and no indirect access is required.
In the odd time step a partial vectorization is performed, which can
avoid the indirect addressing and allows for vectorized execution of
consecutively stored chunks of PDFs.  Nodes that cannot be treated in
this way are updated in the usual way without SIMD vectorization (i.\,e., in
scalar mode).  The fraction of nodes that can be updated with SIMD
operations
depends on the geometry used for the simulation.  During even time
steps, $2 \times 19 \times 8 \text{~(PDF)}$\,\bytes\ $= 304$\,\bytes\ per
\FLUP\ are required.  In the odd time step the number of bytes needed for
one \FLUP\ depends on the fraction of vectorizable updates.  The lower bound
occurs when all updates can be vectorized. In this case only 
$2\times 19 \times 8$\,\bytes\ $= 304$\,\bytes\ of memory traffic are 
required.  The upper bound is reached when all updates must be scalar and
indirect accesses are needed for the adjacency information, 
resulting in $2 \times 19 \times 8
\text{~(PDF)} + 18 \times 4 \text{~(IDX)}$\,\bytes\ $= 376$\,\bytes\
of traffic.

All performance-critical parts were implemented using SIMD compiler
intrinsics to have full control over the code vectorization.  We
restrict ourselves to the SIMD instruction sets SSE4.2
(Streaming SIMD Extensions) and AVX (Advanced Vector Extensions),
which are available on modern x86-based processors.

\subsection{Test bed and benchmark cases}

\supermuc\ \cite{supermuc}, which is installed at Leibniz Supercomputing Center
(LRZ)\footnote{\url{http://www.lrz.de/english/}} in Garching near
Munich, is a tier-0 PRACE\footnote{\url{http://www.prace-ri.eu/}} 
system and one of the main federal compute
resources in Germany. It is built from a number of 512-node
``islands,'' with a fully non-blocking fat tree FDR10 InfiniBand
connectivity inside each island.  A compute node comprises two Intel
Sandy Bridge (``SNB'') EP (Xeon E5-2680) eight-core processors with a base clock
frequency of 2.7\,\GHZ. The actual clock speed of the processors can
be set at job submission time, but the ``turbo mode'' feature of the
Intel processors cannot be used. Consequently, all single-node benchmark tests
were run on a standalone Sandy Bridge EP node with the same type of CPU
and otherwise similar characteristics. Due to the ccNUMA memory architecture,
the memory bandwidth scales perfectly from one to two sockets in a node.

The Intel Fortran/C compiler 13.1 and Intel MPI 4.1 were used in all
cases. The operating system on \supermuc\ was SuSE SLES11.
Each MPI process was explicitly pinned to its physical core
using \verb|sched_setaffinity()| within the code.
All benchmarks were run inside a single island to guarantee that
communication is performed through the fully non-blocking fat tree.

Two geometries were selected for the benchmarks in this paper.
The first is an empty channel which consists only of fluid nodes except 
for the walls.
The second geometry is a \textit{packed bed reactor,} i.\,e., a tube filled with spheres.
It represents a real-world application case for flow simulation with this type
of code.
Both geometries have dimensions of $4000 \times 80 \times 80$ nodes,
resulting in $25 \cdot 10^6$ fluid nodes ($\approx 3.8$\,\GB\ lattice $ +\,1.8$\,\GB\ adj.\,list)
and $19 \cdot 10^6$ fluid nodes ($\approx 2.9$\,\GB\ lattice $+\,1.4$\,\GB\ adj.\,list)
for the channel and the reactor geometry, respectively.

With these dimensions both geometries fit into the NUMA locality 
domain of one socket on \supermuc.
For strong scaling runs the reactor geometry was enlarged to
$8000 \times 160 \times 160$ nodes, as the smaller lattice fits
in the L3 caches of $128$ compute nodes and above.
The large reactor geometry consists of $157 \cdot 10^6$ fluid nodes and requires
around $24$\,\GB\ of memory for the lattice and around $11$\,\GB\ 
for the adjacency list.

The ILBDC code is purely MPI-parallel. All single-node measurements
were thus performed with intra-node MPI only; a hybrid MPI/OpenMP
version exists, but the details of the multi-threaded implementation
(alignment constraints and loop peeling, encoding of obstacle-free
regions for SIMD execution in the odd time step) are beyond the scope
of this work.  We also expect no major differences for an equivalent
OpenMP-parallel version. Details about the MPI parallelization can be
found in Sect.~\ref{sec:mpi}.

\subsection{Contribution}

This paper makes the following contributions:

The scalar and AVX-vectorized single-core performance and
  intra-chip saturation of an LBM implementation with AA propagation
  pattern are successfully described using the ECM model and compared to
  the popular ``pull-split'' propagation pattern. We show the superiority of
  the AA pattern in terms of performance and demonstrate that 
  ``best possible'' performance is achieved on the chip.

The energy consumption of the LBM algorithm with AA propagation
  pattern is modeled on the chip level for a range of clock
  frequencies.  Together with the ECM performance model, we gain a coherent picture
  of the performance and power properties of the LBM algorithm on the
  chip and achieve good qualitative agreement with measurements. A
  region of optimal operating points with respect to clock speed and number of
  cores is identified.  The system's baseline power (power consumption
  of everything apart from the CPUs, i.\,e., memory, chip sets, network,
  disks, etc.)  is taken into account and shown to have a damping
  influence on the differences in energy consumption (as predicted by
  the power model). Even then, potential energy savings of up to 
  50\% can be achieved compared to a naive operating point with
  the inferior pull-split propagation model.
Single-thread code performance and the selection of an optimal
  number of cores per chip (the latter depending on the former)
  are shown to have the largest impact on energy consumption.

In highly parallel LBM runs, we observe a loss in parallel
  efficiency when the CPU clock speed is reduced. This unexpected
  result can be explained by a strong dependence of effective
  inter-node and intra-node MPI communication bandwidth on the clock
  speed. The effective bandwidth also shows a strong negative
  correlation with the number of MPI processes per node.
As a consequence, minimal energy to solution in the highly parallel
  case depends even more strongly on the proper choice of the
  operating point, especially on the number of cores per chip
  (and thus the single-thread performance).
  A simple non-reflective reduction of the clock speed will reduce
  performance and may consume more energy at the same time.

\section{ECM performance model}\label{sec:ecm}

The ECM model \cite{ppam09,hager:cpe13} is a refinement of the
roof\/line model \cite{schoenauer00,williams-2009}.
Note that the ECM model is not specific to the lattice-Boltzmann
algorithm. It has also been used successfully to describe the
performance and scaling properties of streaming benchmark kernels
\cite{ppam09,hager:cpe13}, stencil smoothers \cite{ppl10},
and medical image reconstruction kernels \cite{rabbitct}.  It is to
our knowledge the first approach that can successfully model the
single-thread performance and on-chip scalability of data streaming
applications on multicore processors. 

It starts with an analysis of the serial loop instruction code to get
an estimate for the ``applicable peak performance,'' i.\,e., the
expected maximum performance when all data comes from the L1
cache. After that, all required data transfers needed to get the data
in and out of L1 are accounted for, together with the time it takes to
move the cache lines up and down through the memory hierarchy
levels. Pure streaming is assumed, i.\,e., there are no latency effects
and the prefetching mechanisms work perfectly. Since it is sometimes
hard to predict which of the above contributions to the serial runtime of
a loop can overlap, the ECM model generates a prediction interval
based on no-overlap and full-overlap assumptions, respectively.

Modeling the in-L1 execution and data transfers both require some knowledge about the
architecture, such as instruction latency and throughput, data path
widths, and the achievable memory bandwidth. If some of this data is
unavailable, suitable microbenchmarks or tools can be
employed. Especially for the in-core performance analysis the ``Intel
Architecture Code Analyzer'' (IACA) \cite{iaca} is instrumental in getting a
better view of the relevant bottlenecks. All data transfers between
the memory hierarchy levels occur in packets of one cache line; hence,
the ECM model always considers a ``unit of work'' of one cache line's
length. If, e.\,g., a loop accesses single precision floating-point
arrays with unit stride on a CPU with 64-\byte\ cache lines, the unit
of work is sixteen iterations.

In order to predict the scaling behavior across the cores of
a multi-core chip it is assumed that each core creates some bandwidth
pressure on the relevant bottleneck, which is usually a shared cache
or the chip's memory interface. When the aggregate bandwidth pressure
exceeds the available bandwidth, saturation sets in \cite{suleman:2008}:
\bq
P(t)=\min\left(nP_0,P_\mathrm{max}\right)\eos
\eq
Here, $n$ is the number of cores used, 
$P_0$ is the single-core performance as predicted by
the ECM model, and $P_\mathrm{max}$ is the saturated full-chip performance
for the relevant bottleneck (e.\,g., the memory bandwidth $b_\mathrm{S}$),
\bq
P_\mathrm{max}=I\cdot b_\mathrm{S}\cma
\eq
with $I$ being the computational intensity, i.\,e., how much ``work''
(\flops, \FLUPs, \ldots) is done per byte transferred over the 
bottleneck.  

\subsection{Observed chip-level performance and scaling}

\begin{figure}[tb]
\centering
\includegraphics*[width=0.8\linewidth]{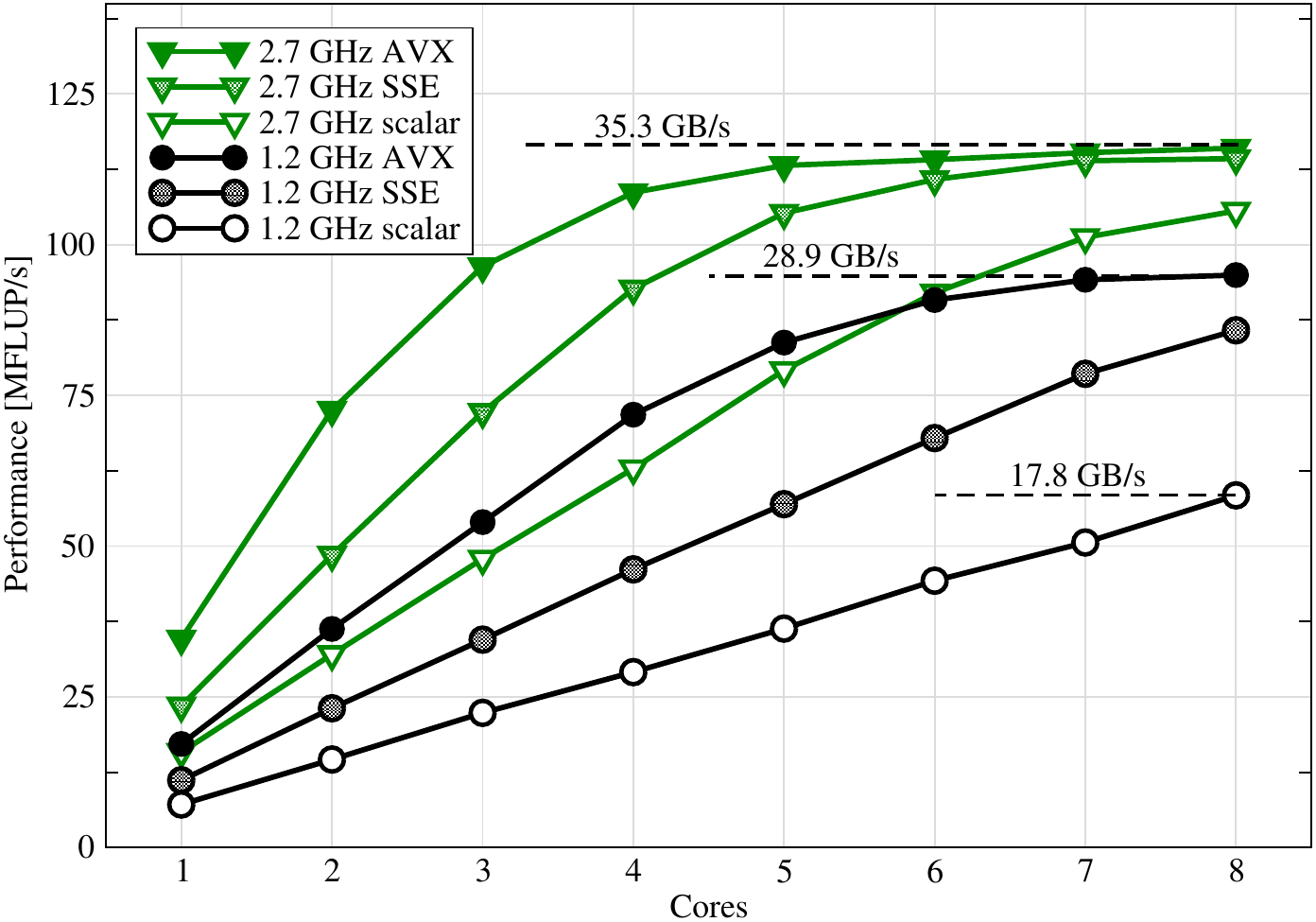}
\caption{\label{fig:AA_vect}Intra-socket strong scaling of the AA pattern
	LBM implementation for an empty channel, comparing AVX, SSE,
	and scalar code at the clock frequencies of
	2.7\,\GHZ\ (triangles) and 1.2\,\GHZ\ (circles). The
	corresponding saturated memory bandwidths are indicated for selected
	cases.}
\end{figure}

As a motivation for doing this kind of analysis, in
Fig.~\ref{fig:AA_vect} we show the intra-socket scaling of the
empty channel test case with the AA pattern 
for the two ``extremal'' clock frequencies 
of 2.7\,\GHZ\ and 1.2\,\GHZ, respectively,  in three variants: 
AVX-vectorized (full 256-bit loads/stores), SSE-vectorized, and 
scalar. 
The data indicates that SIMD vectorization has a large impact
in the serial case; in fact, the serial performance differs
by more than a factor of two between the AVX and the scalar 
code (triangles vs.\ circles). 
At 2.7\,\GHZ, the gap closes as the number of cores is
increased. On the full socket the scalar code is hardly 10\%
slower than the AVX variant. The latter, however, reaches the
same level already with four cores, which opens an opportunity
for saving energy by leaving cores idle. 

At 1.2\,\GHZ, the situation in the serial case is similar, but on a
lower level. The single-core performance of all code variants is
roughly proportional to clock speed. However, only the
AVX-vectorized code shows a saturation pattern, while the SSE and
scalar variants scale linearly up to eight cores without reaching a
bandwidth barrier. This is because these code versions
do not exert sufficient ``pressure'' on the memory interface
to reach saturation even on a full socket.
Hence, lack of vectorization (``slow code'') cannot
be compensated by using more cores in this case.
Moreover, the maximum memory bandwidth is correlated with
the core clock frequency and varies by about 20\% across the
full frequency range~\cite{schoene-2012}. 

\begin{figure}[tb]
\centering
\includegraphics*[width=0.8\linewidth]{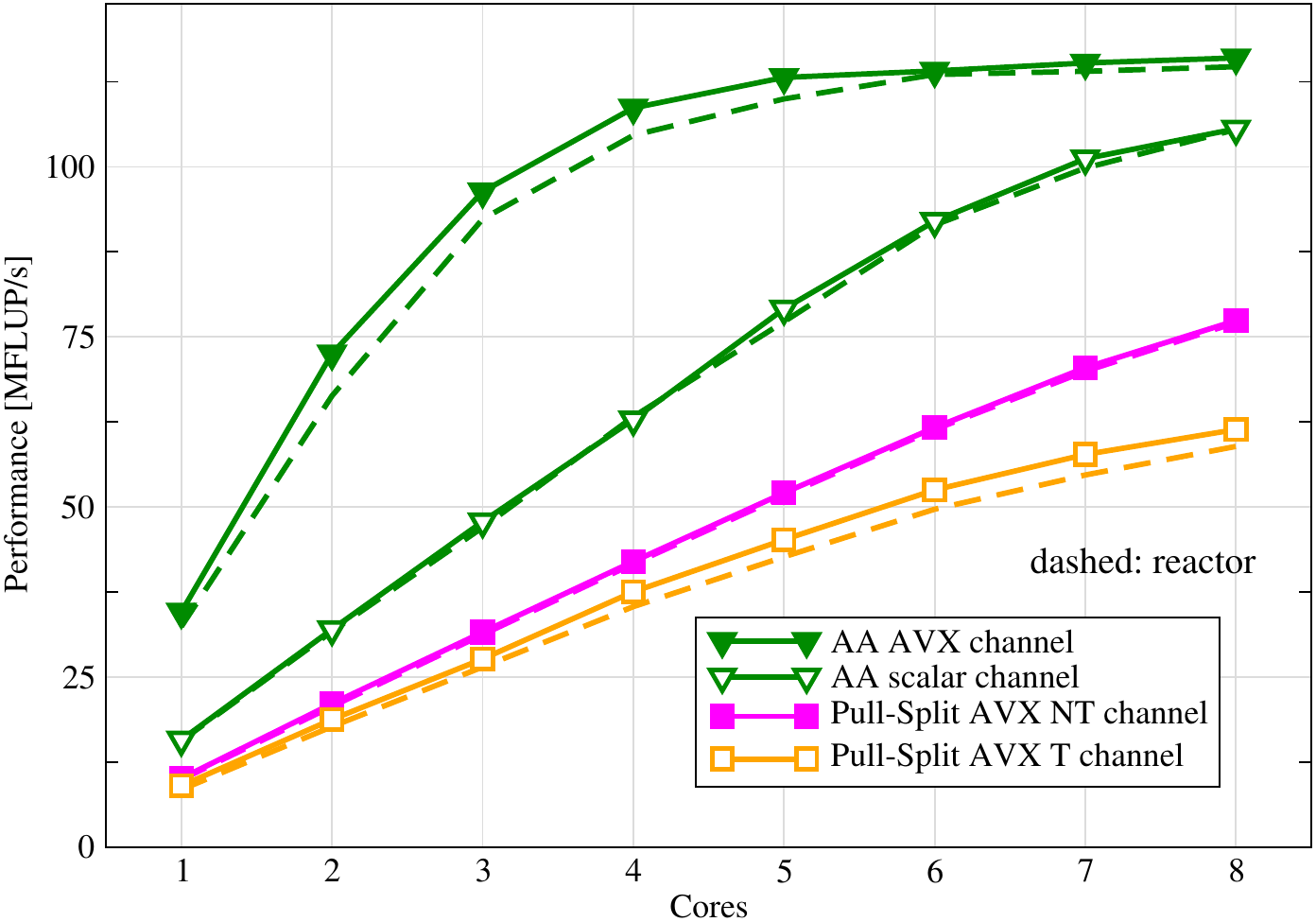}
\caption{\label{fig:socket_channelvspacking}Intra-socket
	performance scaling for one SNB chip at 2.7\,\GHZ: 
	AA pattern in AVX and scalar variants
	(triangles) and pull-split pattern with AVX
	vectorization with (NT) and without (T) non-temporal stores (squares),
        for the empty channel application case (solid lines). The 
	performance numbers for the packed reactor case are shown with
	dashed lines.}
\end{figure}
In Fig.~\ref{fig:socket_channelvspacking} we show a socket-level
performance comparison of the scalar and vectorized AA pattern
implementation with the pull-split pattern for both application cases
(empty channel vs.\ packed reactor) at a clock speed of 2.7\,\GHZ. Although
there is a large fraction of obstacles in the packed reactor geometry, their
presence hardly influences the performance, independent of the
propagation pattern. We also see that the pull-split pattern is not
competitive since it is about half as fast as the scalar AA 
version on the single core, and there are not enough cores available
to compensate for this disadvantage and reach bandwidth saturation.

The intention of applying the ECM model is to gain deeper insight into
this performance behavior and to pave the way for a practically
useful energy consumption analysis.

\subsection{In-core analysis}

An IACA throughput analysis for the AA pattern kernel shows that the
ADD port of the SNB core is the sole bottleneck of core execution for
all variants (scalar, SSE, AVX), as well as for even and odd time
steps, and that one loop iteration (four updates with AVX, two with SSE, one for scalar) 
should take about 135 cycles. In contrast, a critical path
analysis reports somewhat longer execution times due to dependencies
in the instruction and data flow. The critical path depends on the
type of time step. It has a maximum length of 163 cycles (even, AVX),
212 cycles (odd, AVX), 160 cycles (even, scalar), and 187 cycles 
(odd, scalar). This prediction roughly
coincides with direct measurements, which we will use as an input in
the following (160, 212, 158, and 160 cycles, respectively). 
\begin{figure}[b]
\centering
\includegraphics*[width=0.8\linewidth]{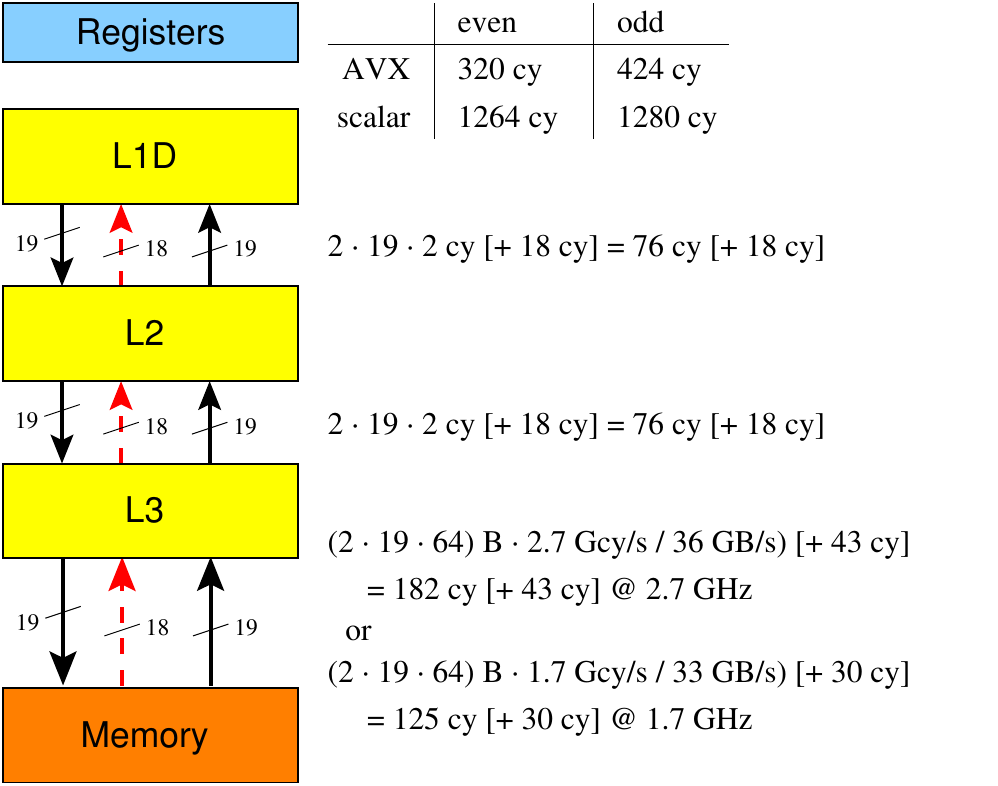}
\caption{\label{fig:AA_model}Single-core ECM model of the AA
  propagation pattern for D3Q19 LBM (eight {\FLUP}s).  Even and odd
  time steps have different in-core timings. One arrow represents the
  number of full cache line transfers indicated; dashed arrows stand
  for half-wide (32-\byte) transfers and are required for loading the
  adjacency information in the odd time step when vectorization is not
  possible.  One half-wide cache line transfer takes one cycle. Numbers in 
  square brackets denote contributions from the adjacency list and can
  be ignored for the empty channel case.}
\end{figure}
These numbers must be multiplied by two (for AVX) or eight (for scalar) 
for getting execution times for one unit of work, i.\,e., a cache
line. The table in Fig.~\ref{fig:AA_model} shows the in-core cycle counts
for AVX and scalar code in the even and odd time steps, respectively.

The
analysis for the packed reactor case is surprisingly very similar: The
even time step does not change at all, since no index access is
required. In the odd time step, even when assuming no potential for
vectorization (as would be the case for an extremely porous geometry)
there is ample room for hiding the additional loads for the index
array due to the bottleneck on the ADD port. This step is necessarily
scalar, however, so the execution time is about 4 times longer per
unit of work. The actual impact of this slowdown depends on the fraction
of vectorizable updates. In our applications, this fraction
is roughly 97\% for the empty channel and 92\% for the packed reactor
case. This leads to a very small performance penalty for the latter,
which was already observed in Fig. \ref{fig:socket_channelvspacking}.
Hence, we will only consider the empty channel case for the rest of
the chip-level analysis.

\subsection{Data transfers and saturation behavior on the chip}

The ECM model requires the maximum attainable memory bandwidth as an
input parameter. It is known that this value depends on the number of
parallel read/write streams as well as the CPU clock speed. From a
data transfer perspective, the AA-pattern implementation of the D3Q19
LBM algorithm reads 19 arrays from memory, modifies their contents,
and writes them back. For measuring the maximum achievable memory bandwidth on
the socket, we therefore use a parallel multi-stream array update benchmark (see
Listing \ref{lst:msu}).
\begin{lstlisting}[caption={Parallel multi-stream update benchmark with 19 streams. \texttt{N} is chosen such that the arrays do not fit in any cache. The runtime is longer than any thread creation or synchronization overhead, so this loop probes the achievable memory bandwidth independently of the programming model.},label=lst:msu,float=t,numbers=none,numberstyle=\tiny]
double a01[N], a02[N],..., a19[N], s=2.0;
#pragma omp parallel for
 for(int i=0; i<N; ++i) {
   a01[i] = s * a01[i];
   a02[i] = s * a02[i];
      ...
   a19[i] = s * a19[i];
 }
\end{lstlisting}
It is designed to mimic the data streaming behavior of the LBM
algorithm.  Care is taken to ensure that the compiler generates the
``intended'' assembly code, i.\,e., full-width aligned AVX loads/stores
are used and loop splitting is avoided. Note that the benchmark in
Listing~\ref{lst:msu} uses OpenMP, but our LBM code is MPI-only. 
This difference is insignificant here because the benchmark serves
as a pure bandwidth probe: any OpenMP-related overhead, such as 
thread creation and barrier synchronization, is orders of magnitude
smaller than the runtime of the loop.
 
\begin{figure}[tb]
\centering
\includegraphics*[width=0.8\linewidth]{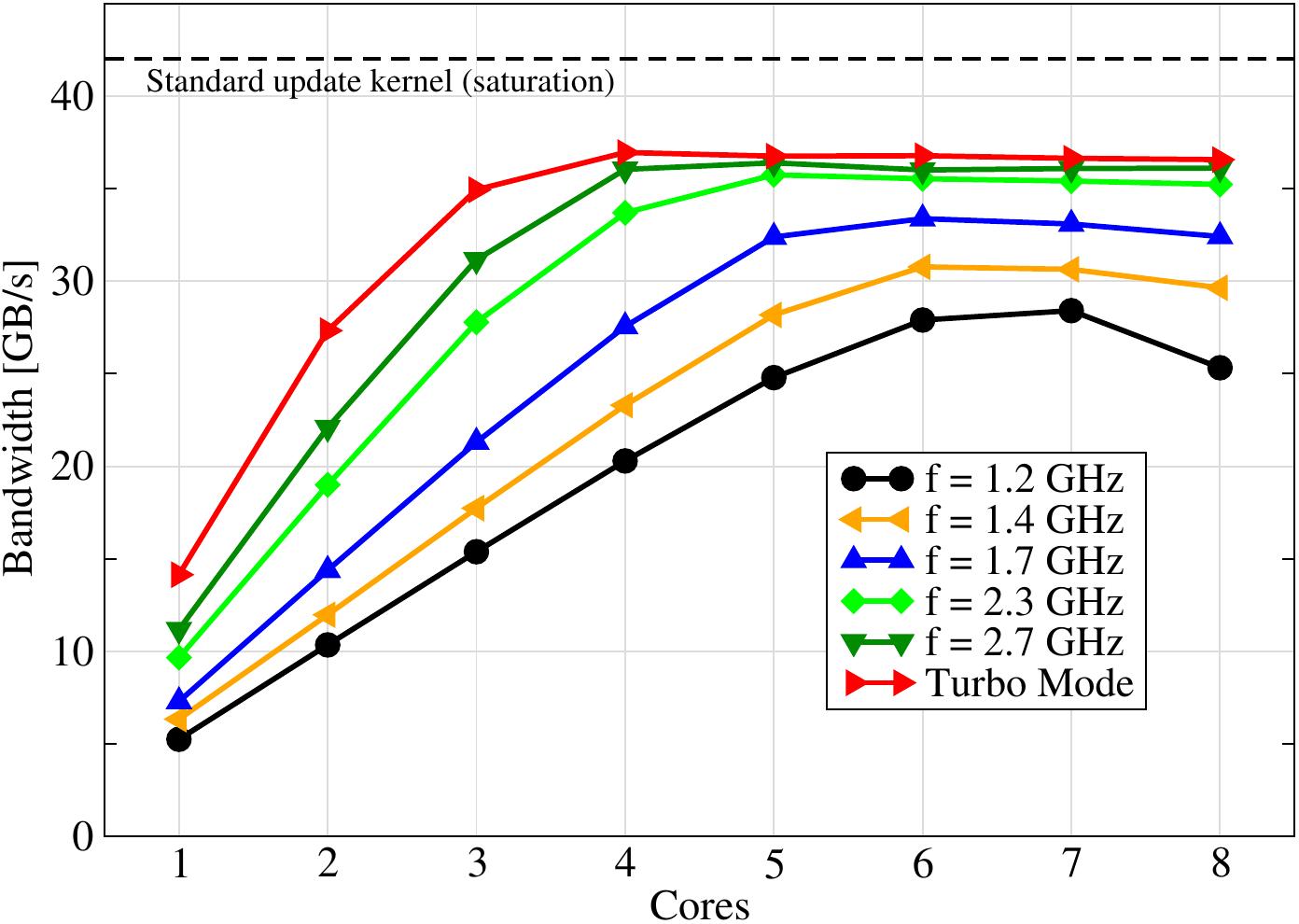}
\caption{\label{fig:MultiStream_socket}Multi-stream update benchmark
	performance scaling on one SNB socket with different CPU
	frequency settings. 19 update streams were run per thread.
	The dashed line indicates the maximum achievable bandwidth
	with a simple single-array update kernel.}
\end{figure}

Figure~\ref{fig:MultiStream_socket} shows the achieved memory
bandwidth on one SNB socket with varying number of threads (cores) and
clock frequencies between 1.2\,\GHZ\ and 2.7\,\GHZ\ (plus turbo
mode). As predicted by the ECM model, the single-thread performance is
proportional to the clock speed and the saturation point is shifted
to larger thread counts as the clock speed decreases: while saturation
is reached near three cores with turbo mode, up to six cores are needed
at the lowest frequencies. 
Due to the large number of read/write streams, the maximum bandwidth
is significantly lower than with a standard single-stream update
kernel (dashed line in Fig.~\ref{fig:MultiStream_socket}). 
At the same
time, the maximum (saturation) memory bandwidth drops by about 25\%
over the whole frequency range; there is another substantial drop
when using the full socket (eight cores) at the lowest frequency.  
As of now we have no conclusive
explanation for these latter effects. They do, however, influence
the considerations on energy dissipation, which will be discussed
in Sect.~\ref{sec:power}.
In the following we will use the maximum bandwidths as 
measured at the respective frequencies as an input to the ECM model
in order to calculate the number of cycles required to transfer 
cache lines between memory and L3 cache.  

Fig.~\ref{fig:AA_model} shows the complete ECM model analysis at
2.7 and 1.7\,\GHZ, respectively. The cycle counts in square brackets
are contributions from loading the adjacency information (dashed
arrows), and can be ignored for the empty channel case. The achievable
memory bandwidth (36\,\GBS\ and 33\,\GBS, respectively) and the clock speed
enter the model when calculating the cycles for data transfers
to and from main memory. Data transfers between
adjacent cache levels are assumed occur at 32\,\bytes\ per cycle,
so these cycle counts are independent of the clock frequency.
The various execution and data transfer times may be combined in 
different ways to arrive at a performance prediction for the serial 
program:
\begin{enumerate}
\item The most
conservative (worst case) assumption is that none of those contributions overlap
with each other, so that the execution time is equal to their sum
(e.\,g., 320+2$\cdot$76+182=654 cycles for the even time step with AVX 
at 2.7\,\GHZ).
\item The most optimistic assumption is that the cycles in
which the L1 cache is occupied by loads and stores from the core
cannot be used for reloads and evicts to L2, but all other
contributions do overlap.
\item Lastly one may assume that the pure in-core execution part 
(everything except loads and stores) can overlap with loads and evicts
from/to the L2 cache, but that there is no overlap beyond that.
\end{enumerate}
\begin{figure}[tb]
\centering
\includegraphics*[width=0.8\linewidth]{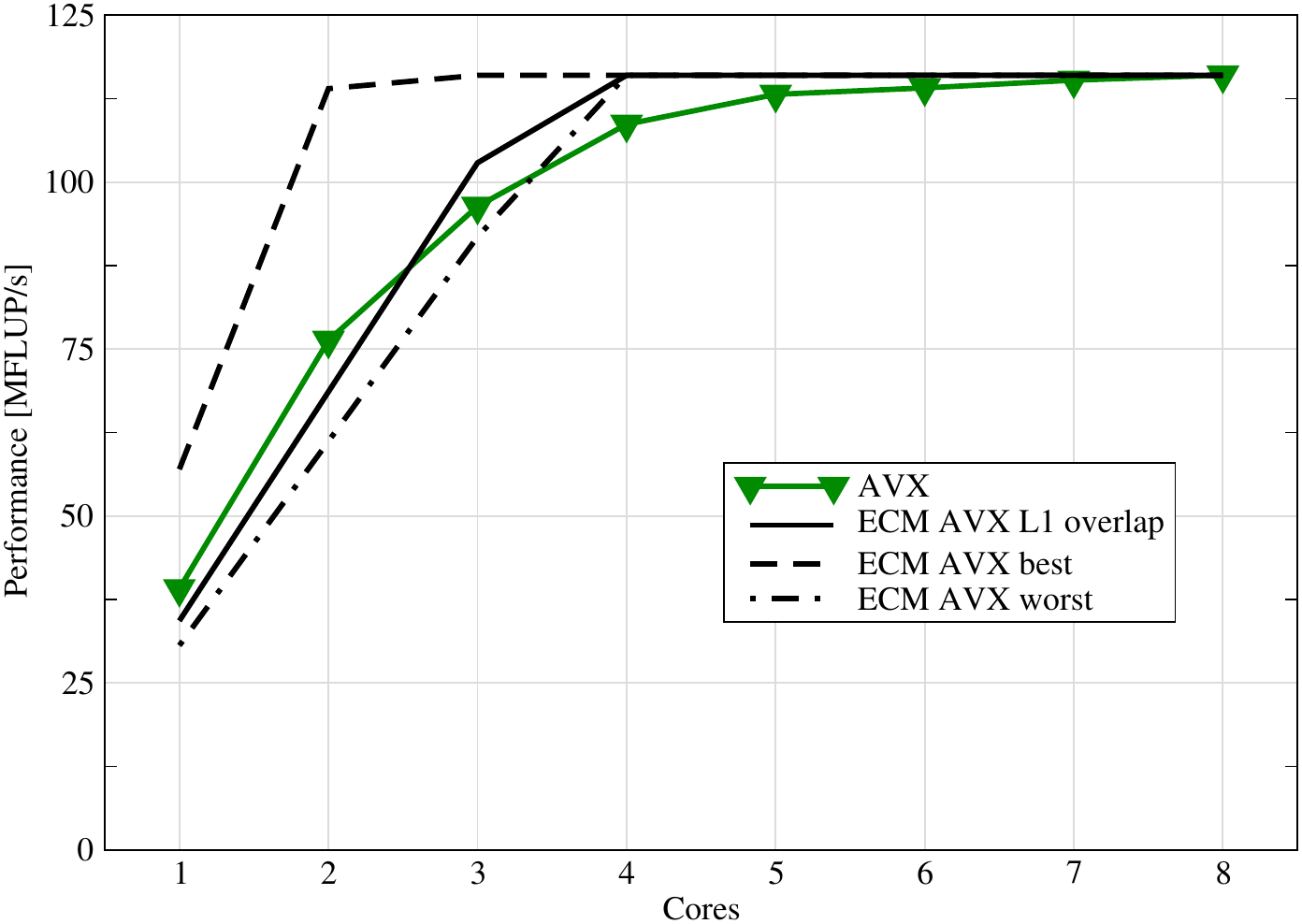}
\caption{\label{fig:ECM_AA_perf}Performance of the AVX (triangles)
  implementation of the LBM algorithm with AA propagation (empty
  channel case) at 2.7\,\GHZ. The ECM model predictions for AVX with
  full overlap assumption (dashed line), no overlap (dotted-dashed line), and
  partial overlap at L1 (solid line) are shown for comparison. 
}
\end{figure}
None of these assumptions coincides with the roof\/line model, which
requires the achievable memory bandwidth for each number of
cores as an input parameter. The ECM model only requires
the maximum (saturated) bandwidth and it predicts the scaling.

\subsection{Validation of the performance model}

Figure \ref{fig:ECM_AA_perf} shows a comparison of the measured
performance for the AVX-vectorized AA pattern implementation
with the three models described above. Apart from the
region around the saturation point (3--4 cores), the third assumption
provides the best fit to the data. 

It was already shown in Fig.~\ref{fig:socket_channelvspacking} that
the pull-split propagation pattern (with and without NT stores) is not
competitive since it cannot saturate the memory bandwidth, although
the NT version has almost the same computational intensity as the AA
pattern. This failure can mainly be attributed to the fact that the
pull-split variant cannot be efficiently SIMD-vectorized on the Sandy
Bridge architecture due to the indirect access in every lattice site
update. More specifically, the loop which loads the neighboring
distribution functions and stores intermediate results into temporary
buffers is scalar. We will thus ignore pull-split from now on and
focus the following discussion on the AA pattern.


\section{Power model}\label{sec:power}

Recently, considerations of power dissipation and energy to solution
have received much interest in the supercomputing community. The two
prevalent questions are: (i) ``How can a parallel code be run so that
its overall energy consumption until a solution is reached can be
minimized, preferably under the constraint of constant time to
solution?'' and (ii) ``How can a parallel computer be operated in a
production environment so that overall power dissipation is minimized
or kept below a given maximum?''

We concentrate on the first question here.
In \cite{hager:cpe13} we have established a simple power model
which is able to explain many of the peculiar properties of the
energy-to-solution metric on a multicore processor for load-balanced
codes that may show some performance saturation as the number of
cores used is increased. In the following, we briefly summarize
its derivation and the most important conclusions drawn from it.

\subsection{Power versus clock speed and energy to solution}

Direct measurements on the Sandy Bridge chip \cite{hager:cpe13} with
the RAPL interface~\cite{intel-sdm-2013-03,10.1109/MM.2012.12}
and the \texttt{likwid-powermeter} 
tool\footnote{\url{http://code.google.com/p/likwid}} show
an expected strong growth of power dissipation 
with rising clock frequency. Although it is not clear what the actual
dependence should be, measurements suggest a power law with an
exponent between 1 and 3~\cite{hager:cpe13}. This is presumably
due to a hard-wired, chip-specific mapping of the clock speed setting to the
core supply voltage. Motivated by measurements on a variety 
of current Intel processor models, we assume a quadratic
behavior so that the power dissipation is
\bq
W(f,n)=W_0 + \left(W_1f +W_2f^2\right)n\cma
\eq
where $f$ is the clock frequency and $n$ is the number of active cores.
The part which is linear in $n$ is the dynamic power dissipation,
whereas $W_0$ is the baseline power. $W_0$ may include contributions
from the rest of the system, such as memory, I/O circuitry, network,
disks, and cooling. The parameters $W_1$ and $W_2$ characterize the
interaction of the code with the hardware in terms of energy consumption,
and they depend on the code being executed and on the data transfers through
the cache hierarchy. They vary by no more than a factor of two 
even across very different code characteristics on the CPU 
considered here~\cite{hager:cpe13} and can be determined by direct
measurements at different clock speeds. Here we choose 
$W_2\approx 1\,\W/\GHZ^2$, $W_0\approx 23\,\W$ for the chip level, 
and $W_0\approx 73\,\W$
for the per-socket share of the whole system. 
Note that we do not expect these
values to be exact (they even vary across multiple specimens 
of the same processor type), but they are sufficient for a qualitative
analysis. Since $W_1$ is  very small on the processors considered here, 
we neglect it in the following.\footnote{CPUs with a low nominal clock
speed (e.g., Intel Ivy Bridge at 2.2\,\GHZ) tend to show a more 
linear characteristic, so that $W_1\gg W_2$. }

Energy to solution is proportional to the ratio of power dissipation
to performance (for a constant problem). We assume that
multi-core performance can be modeled by
\bq
P(t) = \min\left(\frac{f}{f_0}\,nP_0,P_\mathrm{max}\right)\cma
\eq
where $f_0$ is the base (``nominal'') frequency of the chip, 
$P_0$ is the serial
performance at $f_0$, $n$ is the number of active cores, 
and $P_\mathrm{max}$ is a maximum performance (see also Sect.~\ref{sec:ecm}).
$P_\mathrm{max}$ may be set by the presence of some bottleneck
such as the memory bandwidth, or it may be infinite if the code
is perfectly scalable. Hence, the energy to solution is
\bq\label{eq:powermodel}
E = \frac{W_0 + W_2f^2n}{\min\left(\frac{f}{f_0}\,nP_0,P_\mathrm{max}\right)}\eos
\eq
The ``roof\/line model of energy'' by Choi et al.~\cite{choi:2012}
also uses direct power measurements and microbenchmarks to fix model
parameters such as the cost for a data transfer or for a
floating-point operation, but it does not address frequency dependence
and multicore scaling, and it reduces the code characteristics to a
single number (computational intensity). These properties make it
very useful for comparisons between architectures and for
design space exploration, while our model is more phenomenological. 
Still the method from \cite{choi:2012} could be used to determine
or at least estimate $W_0$, $W_1$, and $W_2$.

It is a simple consequence from this model that a minimum energy to
solution is reached near the performance saturation point, i.\,e., at the
smallest number of cores $n$ for which $P(n)=P_\mathrm{max}$.  If by some
means $P_\mathrm{max}$ can be increased, e.\,g., by choosing the AA propagation
pattern, energy to solution will
immediately go down proportionally.
In case the available number of cores is too small to get saturation,
i.\,e., if the performance scales well, all cores must be used for minimum
energy consumption.

The single-core performance $P_0$ also appears only in the denominator.
In the application cases we consider here, it is mainly influenced
by the SIMD vectorization and thus, indirectly, by the choice of 
propagation pattern. The larger $P_0$, the fewer cores are needed
to reach saturation, so there is an inherent energy saving potential
in having a fast single-threaded code. 

The dependence of $E$ on the clock frequency is more complex: If there
is no saturation there may be an optimal frequency $f_\mathrm{opt}$
for which $E$ is minimal:
\bq\label{eq:fopt}
f_\mathrm{opt}=\sqrt{\frac{W_0}{W_2 n}}
\eq
If $W_0$ comprises only the chip's idle power, $f_\mathrm{opt}$ is in the range
of accessible frequencies. However, a low frequency setting has the
adverse effect of extending the time to solution, which may not be
desirable.  Cost models other than time or energy (such as the
energy-delay product) may then be needed to determine whether this
is acceptable. On the other hand, if $W_0$ contains (the chip's share
of) the baseline power of the complete node, $f_\mathrm{opt}$ is usually
near or beyond the highest possible setting. We call this 
``clock race to idle,'' because a fast clock speed saves energy. 
The details of this analysis can be found
in \cite{hager:cpe13}.

\subsection{Energy to solution for the LBM solver on the chip}\label{sec:ets_chip}

The ECM model and the power model enable a combined analysis of 
the energy and performance properties of the LBM
algorithm. It is useful to plot energy to solution
versus performance, with the number of cores used as a parameter
within a data set for a specific frequency, SIMD vectorization
variant, propagation method, or any other property. 
\begin{figure}[tb]\centering
\includegraphics*[height=0.6\linewidth]{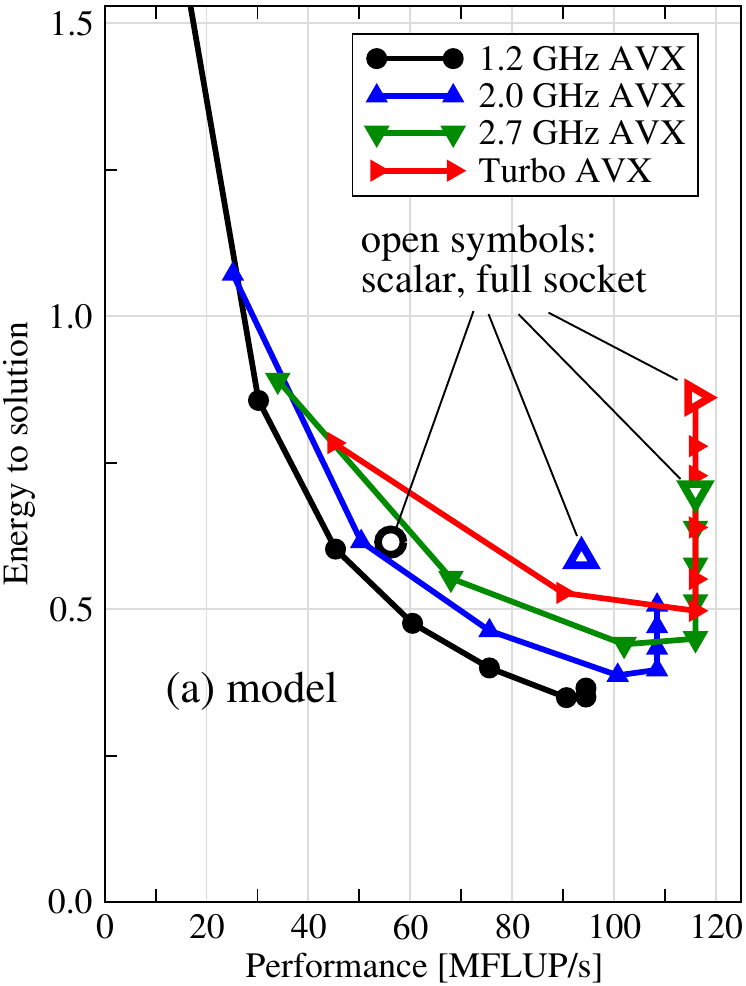}\hspace*{-0.5mm}
\includegraphics*[height=0.599\linewidth]{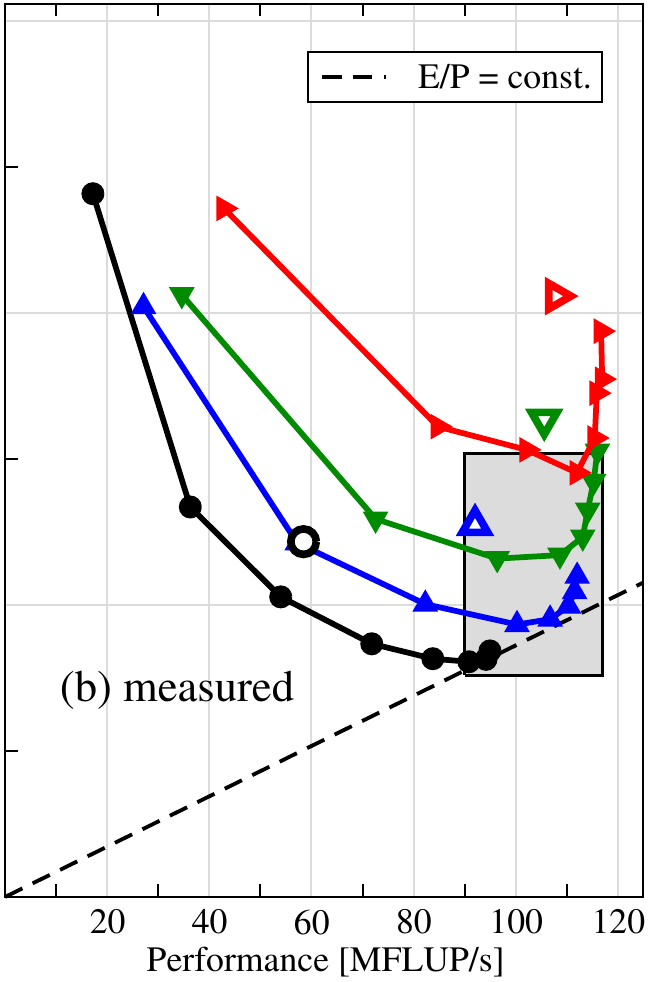}
\caption{\label{fig:Z_ets_AAvsScalar}Energy to solution
  vs.\ performance of the AVX-vectorized LBM AA pattern implementation
  (empty channel case) of one SNB socket for different clock
  frequencies (lines and filled symbols).  The number of cores used 
  (processes per chip, PPC) is
  the parameter along each data set. (a) Predictions by the ECM
  performance model and the chip-level power model. (b) Measured data.
  For comparison, the big open symbols mark the energy and performance
  of the scalar code on a full socket.  The shaded area is the region
  defined by absolute minimum energy and saturated performance for the
  AVX versions. The dashed line is the line of constant energy-delay
  product that hits the saturation point of the lowest-frequency run.
  Note that all legends in (a) also apply to (b).}
\end{figure}
This has been done in Fig.~\ref{fig:Z_ets_AAvsScalar}a for three
different clock frequencies and turbo mode, using the AA pattern in
the AVX variant (filled symbols).  For comparison we also show the
energy-performance data for full socket runs with purely scalar
code (large open symbols). In
turbo mode, each data point was computed using the maximum allowed
frequency for each number of active cores. The corresponding
measurements are shown in Fig~\ref{fig:Z_ets_AAvsScalar}b.
Note that we always show energy to solution in arbitrary units,
but the values shown are coherent for a specific problem size
(geometry and number of iterations).

The models are able to describe the qualitative features of energy and
performance. The observed deviations are caused by (i) the inability
of the ECM model to accurately describe the performance behavior in
the vicinity of the saturation point, (ii) the inaccuracy in
determining $W_2$ and $W_0$, and (iii) the approximation of linear
power behavior with respect to core count even with saturated codes
like LBM at higher clock speeds. In addition, turbo mode does not fit
perfectly into the model (\ref{eq:powermodel}) since the SNB chip can
operate beyond its thermal design power (TDP) for a limited amount of
time \cite{10.1109/MM.2012.12}. This is why the deviation from the
measurements is especially large with turbo mode (right-pointing
triangles in Fig.~\ref{fig:Z_ets_AAvsScalar}).

\begin{figure}[tb]
\centering
\includegraphics*[width=0.8\linewidth]{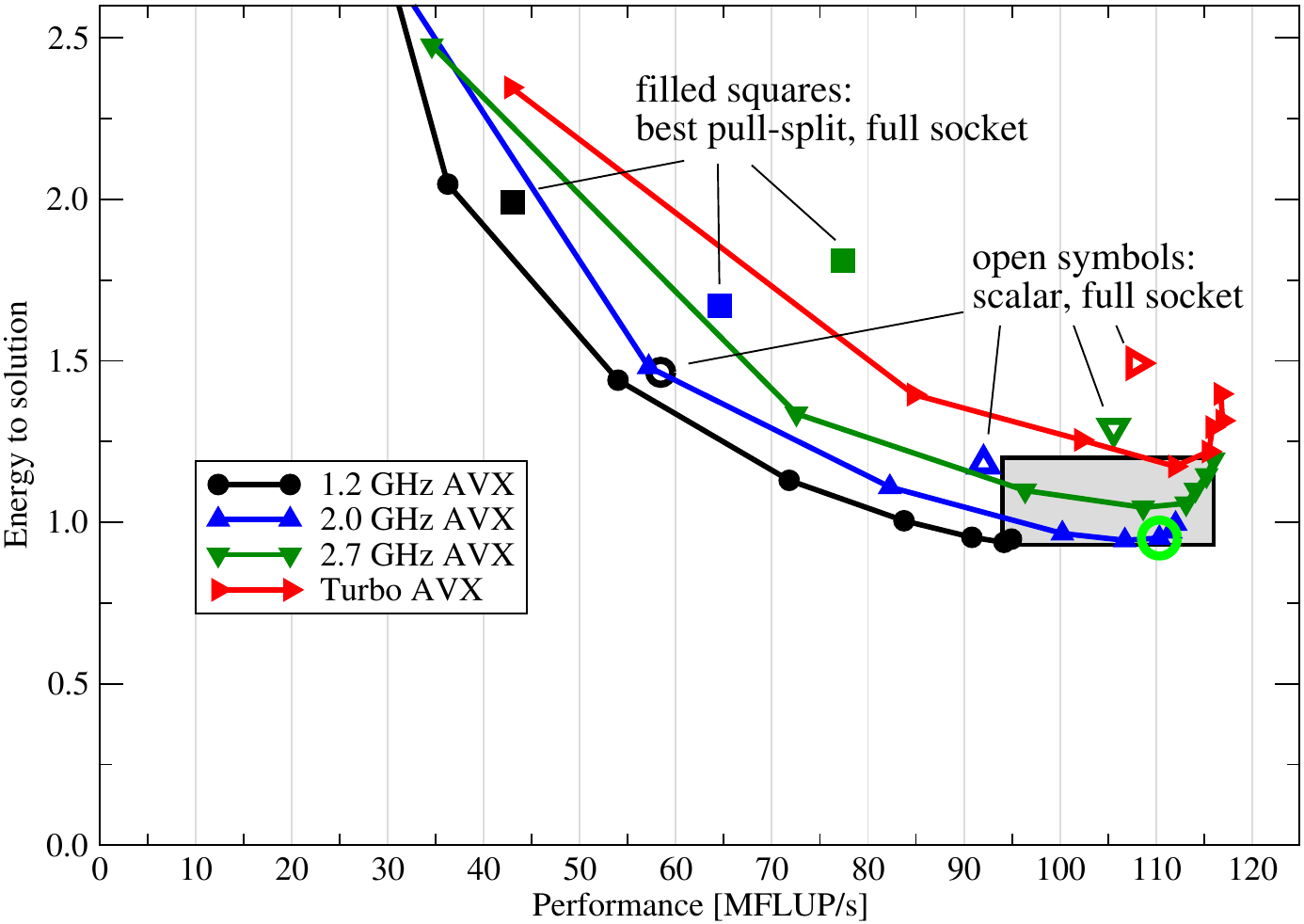}
\caption{\label{fig:Z_ets_AAvsScalar_full}Same data as in 
	Fig.~\ref{fig:Z_ets_AAvsScalar}b but with a realistic
        per-socket baseline
	power of $W_0'=W_0+50\,\W=73\,\W$. The circle marks 
        a possible optimal operating point for almost minimal energy
        with a tolerable loss in performance. For reference, 
        the best data for the pull-split propagation pattern 
        (vectorized, full socket) for 1.2, 2.0, and 2.7\,\GHZ\ 
        is also shown (filled squares).}
\end{figure}
Scalar code (large open symbols) is only able to (almost) saturate the memory 
bandwidth at 2.7\,\GHZ, and it is neither competitive
in the energy nor in the performance dimension. 
Looking at the minimum energy point with respect to clock
frequency and number of cores in the regime where performance is not
saturated, we see that this point moves to smaller frequency as the
core count goes up, as described by Eq.~(\ref{eq:fopt}). In general, 
a faster sequential code (AVX instead of
scalar) saves energy. Comparing energy to solution for the AVX codes
at their respective saturation points, we can identify an
``optimization space'' (shaded area in
Fig. \ref{fig:Z_ets_AAvsScalar}b), in which the desired optimal
operating point should be found. Depending on the emphasis one wants
to put on energy minimization vs.\ maximum performance, this point
may be in the lower left corner of the area. In this case one would use all
cores at the lowest frequency (1.2\,\GHZ, filled circles) and
sacrifice about 20\% of performance compared to the right edge
of the area, which is defined by the saturation point at higher frequencies
(2.0\,\GHZ\ to turbo mode). Another clear conclusion is that turbo mode
is of no good use for the LBM implementations studied here, neither from a 
performance nor from an energy point of view. 

There is no single, well-defined criterion for identifying the optimal
operating point on the chip level. One may certainly employ cost
models such as the energy-delay product (ratio of energy and
performance), but this is only one possible choice. For reference we
have included a line of constant energy-delay product in
Fig. \ref{fig:Z_ets_AAvsScalar}b. From the data we have collected,
using 5--6 cores at 2.0--2.3\,\GHZ\ seems to provide a good compromise
between performance loss and energy consumption (``as far on the lower
right as possible'').

While the model and the measurements yield a consistent picture on the
chip level, it is clear that the chip contributes only a (however
significant) part to the overall power consumption of a compute
node. As mentioned above, the rest of the system should be taken
into account when assessing the real energy demand for
running an application. We do this by adding an extra,
constant contribution of 50\,\W\ to the chip-level baseline power,
so that the new baseline power is $W_0'=23\,\W+50\,\W=73\,\W$.
This leads to roughly 300\,\W\ of maximum node 
power (assuming two-socket nodes and a maximum power dissipation 
per socket of 100\,\W), which is the value
measured during a LINPACK run on \supermuc\ \cite{hubermuc}.
With this change we can offset the energy measurements from
Fig.~\ref{fig:Z_ets_AAvsScalar} to arrive at the data shown in
Fig.~\ref{fig:Z_ets_AAvsScalar_full}. Note that the modified
baseline $W_0'$ also contains everything beyond the chip, especially
including the network and the cooling infrastructure, 
broken down to the socket level. We thus assume that these contributions
are largely constant compared to the dynamic power variations
parametrized by $W_2$, $f$, and $n$ in Eq.~(\ref{eq:powermodel}).

As expected, the modified baseline power leads to a reduction of the
vertical spread between the measurements for different clock
frequencies. While it was possible with the chip-level (i.\,e., small)
$W_0$ to have a situation where energy to solution was heavily
influenced by frequency and SIMD vectorization even at a specific
performance level (with a spread of up to 2$\times$ within the
optimization space shown in Fig.~\ref{fig:Z_ets_AAvsScalar}), the large
$W_0'$ reduces the spread to about 25\%. Hence, a large baseline
power favors the ``race to idle'' principle where the most influential
parameter is performance; optimizations that favor a larger saturation
performance (such as the AA propagation pattern or blocking schemes
which increase the computational intensity) have the most potential
for saving energy. 
In addition, optimized clock speed and a reduction of the number of
cores used can yield second-order but still significant savings.
Within the transformed optimization space (shaded area in
Fig.~\ref{fig:Z_ets_AAvsScalar_full}) we can identify a possible
optimal operating point at about 2.0\,\GHZ\ and six cores, with almost
minimal energy to solution and a performance loss of about 6\%
compared to the highest possible saturation level. In comparison to a
naive strategy of running on all cores with turbo mode enabled and a
scalar kernel, more than one third of the energy can be saved.

The ``race to idle'' principle with respect to maximum code
performance is evident from a comparison with the energy-performance
data for the pull-split pattern (filled squares) in the best variant
(SSE or AVX vectorized, non-temporal stores, full socket) at three
different frequencies in Fig.~\ref{fig:Z_ets_AAvsScalar_full}: 
The pull-split pattern can neither compete with
AA in the performance nor in the energy dimension. Using AA, almost a
factor of two in energy and 30--40\% of runtime can be saved in
comparison to pull-split.


\section{MPI-parallel LBM simulations}\label{sec:parallel}

\subsection{MPI parallelization in ILBDC}\label{sec:mpi}

\begin{figure}[tb]
\centering
\includegraphics*[width=0.95\linewidth]{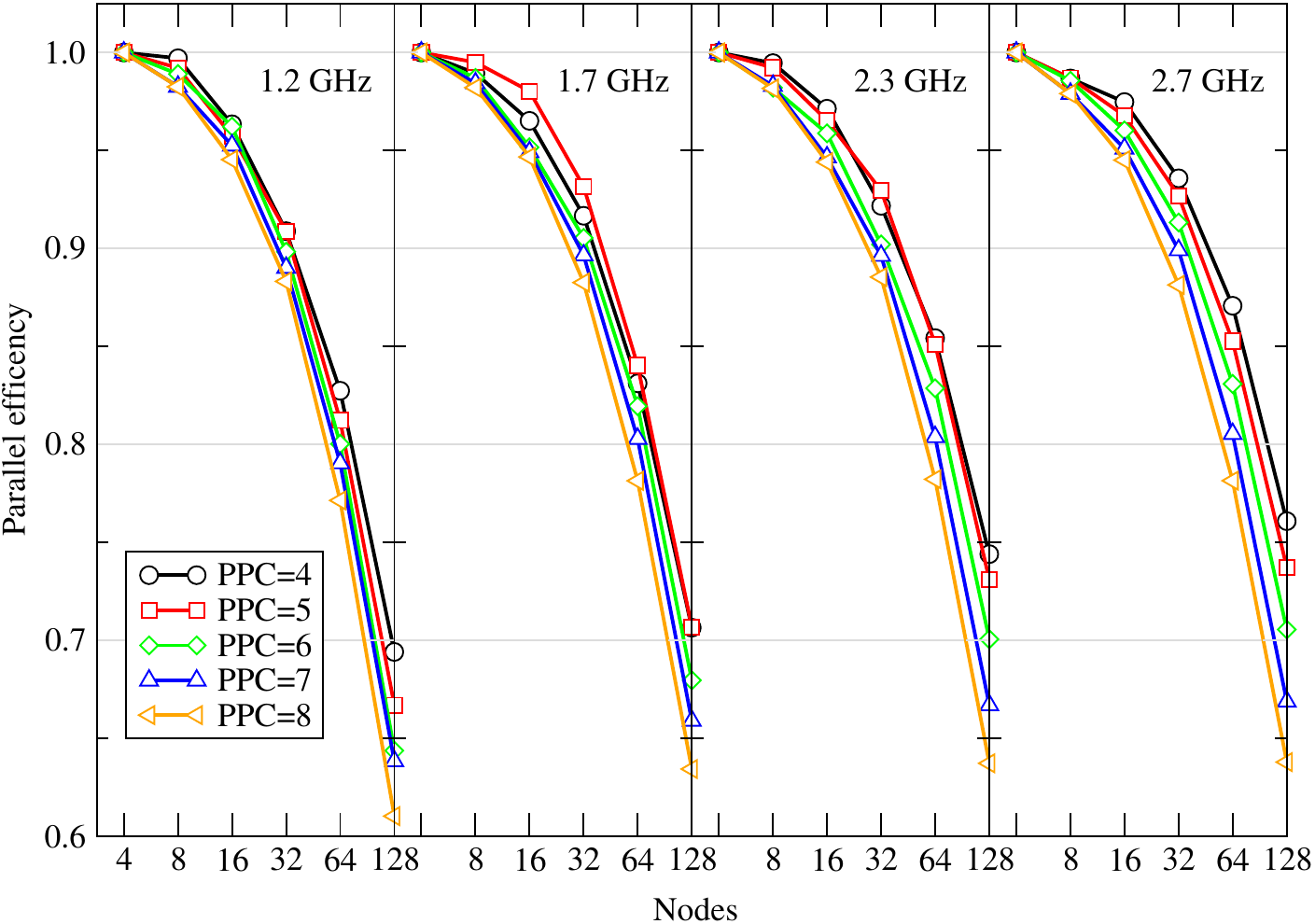}
\caption{\label{fig:par_eff_packing_scaling}Parallel efficiency of
  the large packed bed reactor application case ($8000 \times 160 \times 160$
  lattice nodes) for different frequency settings and different number
  of processes per chip (PPC) on up to 128 nodes of \supermuc. The efficiency
  calculation was based on the four-node performance baseline.}
\end{figure}
ILBDC uses an MPI parallelization with a static load balancing scheme.
The sparse representation of the lattice is cut into equally sized chunks, 
so that each MPI rank receives the same number of fluid nodes.
The interfaces of such generated partitions can be arbitrarily formed 
with different numbers of partition neighbors, as the simple cutting
of the sparse representation does not consider any topological 
information.
However, in the case of the channel and reactor benchmark geometries 
this method results only in a 1-D 
decomposition, where each rank only needs to exchange ghost PDFs with 
its two direct neighbors.
%
We distribute the ranks linearly across the compute nodes, so that consecutive
ranks are located nearby on the same node.
Due to the implicit one-dimensional domain decomposition, 
only the first and the last rank on a compute node 
must then perform inter-node communication. All the remaining ranks
communicate within the node only.
For the same reason, in case of strong scaling the communication
volume of each process stays constant when the number of processes
goes up: each rank gets a successively smaller slice of the long
geometries.

The packed bed reactor geometry was used for all the multi-node
experiments, since it is the application scenario that is
relevant in practice. We have shown earlier that the node-level
performance (and thus power) properties are very similar 
to the empty channel case. Note also that ``turbo mode'' cannot
be activated on \supermuc, so we stick to the fixed frequencies
of 1.2, 1.7, 2.3, and 2.7\,\GHZ\ in the following.

\subsection{Performance and energy at strong scaling}

\subsubsection{Parallel efficiency and communication performance}

All variants of the AA pattern scale well up to 32 nodes (512 cores)
at all frequencies and parallel efficiency only starts to degrade
below 90\% beyond that point. Scaling experiments were performed on
up to 128 nodes (2048 cores), since this is where some variants start
to show efficiencies as low as 60\%. 
In Fig.~\ref{fig:par_eff_packing_scaling} we show the parallel efficiency
of the strong scaling runs versus the number of nodes at the four chosen
frequencies and with between four and eight processors per chip (PPC). 
Since the application case is too large to fit on a single node, 
all efficiency numbers were normalized to the four-node run.

Usually one would expect the parallel efficiency to increase as the
node-level performance goes down, because communication and
synchronization overheads become less important when the pure compute
time goes up. On \supermuc, the opposite is the case: the minimum
parallel efficiency (at 128 nodes) varies between 76 and 63\%
(depending on the number of processes per chip) for 2.7\,\GHZ, but between
69 and 61\% at 1.2\,\GHZ. We conclude that there must be a
frequency-dependent factor which impedes scalability whenever
communication overhead plays a significant role.

\begin{figure}[tb]
\centering
\includegraphics*[width=0.8\linewidth]{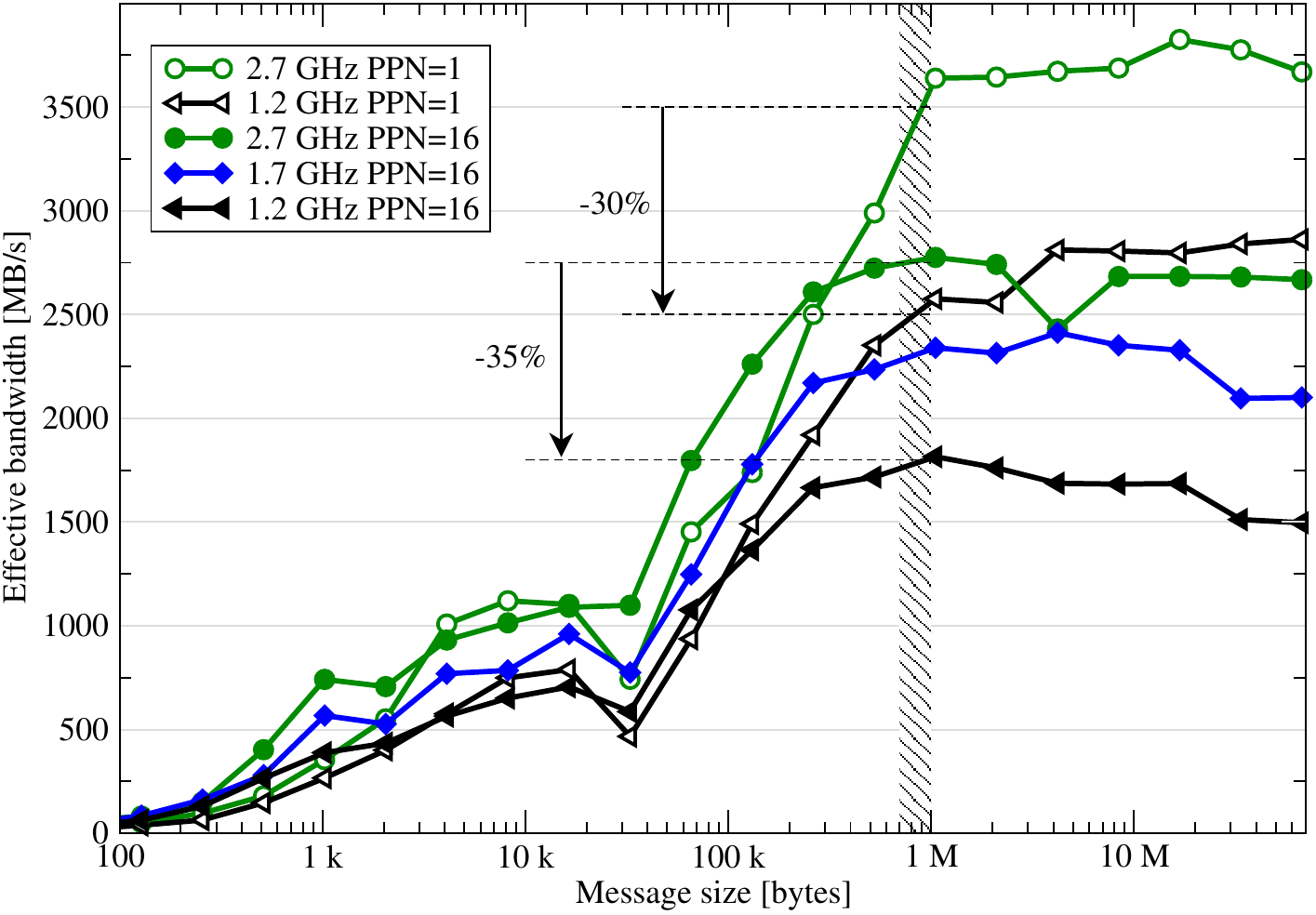}
\caption{\label{fig:sendrecv}IMB sendrecv benchmark results on two
  \supermuc\ nodes for different CPU clock speed settings with 16
  processes per node (filled symbols) and one process per node (open
  symbols). The mapping of MPI ranks to cores was set for minimum
  inter-node traffic (consecutive ranks on the cores of a node). The
  shaded area indicates the range of message sizes for the LBM
  application test case (packed bed reactor geometry). The
  effective communication bandwidth drops by 35\% across
  the available frequency range at PPN=16 and still by 30\% at 
  PPN=1 (arrows).}
\end{figure}

In order to explore the reasons for this effect we have conducted
experiments with ``sendrecv'' from the Intel MPI benchmark suite (IMB)
\cite{imb}, since it mimics the ringshift-like
halo-exchange communication pattern of the ILBDC code. Each MPI
process exchanges data with its neighbors:
\texttt{MPI\_Sendrecv(to right neighbor, from left neighbor)}.
The benchmark reports the available communication bandwidth per
process.  In Fig.~\ref{fig:sendrecv} we show the results for two
\supermuc\ nodes in the two corner cases of one process (PPN=1) and 16
processes per node (PPN=16) for the two extremal frequencies of
1.2 and 2.7\,\GHZ. The placement of the MPI ranks was done
in the same way as for the ILBDC benchmarks: neighboring ranks were
``packed'' to the same node to minimize inter-node traffic.

Although both scenarios show a dependence of the effective MPI
bandwidth on the clock speed, this is especially pronounced at PPN=16,
and we see a breakdown of about 35\% in communication bandwidth
within the region of message sizes relevant for the ILBDC packed reactor
benchmark (shaded area). Moreover, the bandwidth of the FDR-10 IB
interface cannot be saturated even at the highest frequency setting
with PPN=16. We attribute both effects to the dominance of intra-node
communication, which has a strong dependence on clock speed.\footnote{Note that even with PPN=1 (open symbols in Fig.~\ref{fig:sendrecv}) 
there is still a roughly 30\% breakdown in
effective bandwidth, so the observed effect would be visible
even with a single process per node, where intra-node communication
does not take place.} In
contrast, the saturated LBM performance with the AA pattern and AVX
vectorization only drops by about 20\% over the whole frequency
range (see Fig.~\ref{fig:AA_vect}). This explains the stronger
breakdown of parallel efficiency at strong scaling and low clock
speeds.

\subsubsection{Energy and performance at scale}

\begin{figure}[tb]
\centering
\includegraphics*[height=0.53\linewidth]{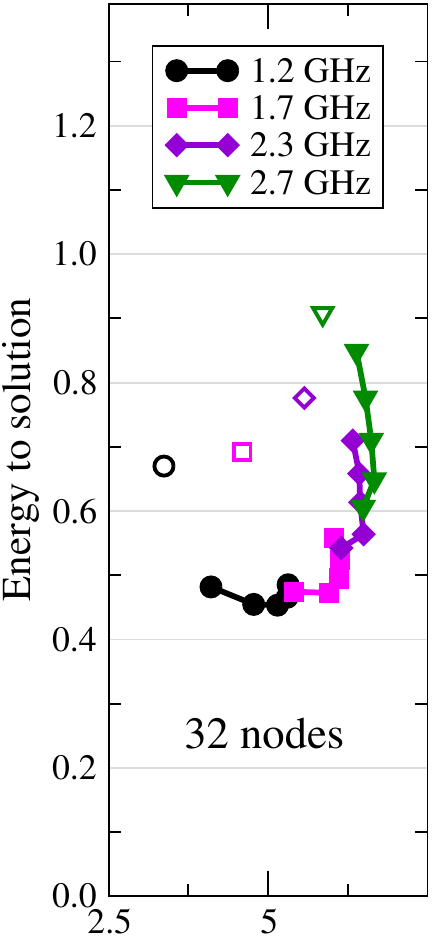}\hfill
\includegraphics*[height=0.53\linewidth]{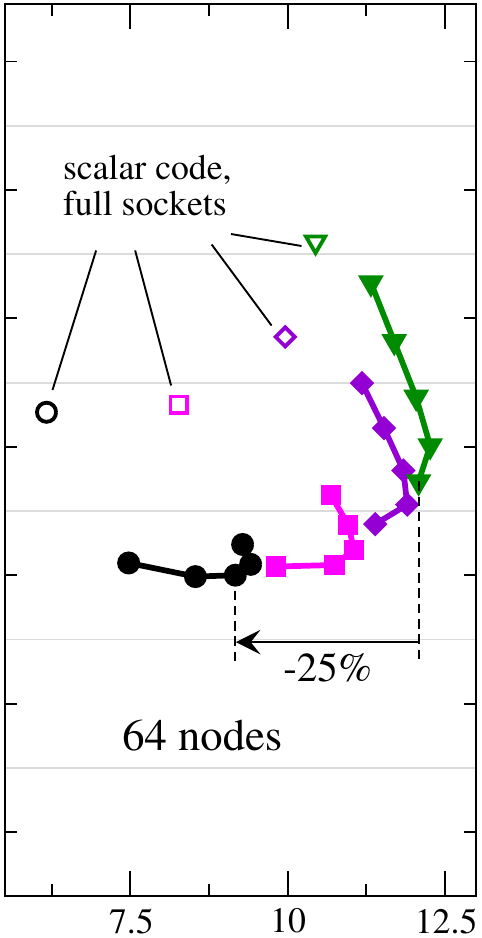}\hfill
\includegraphics*[height=0.53\linewidth]{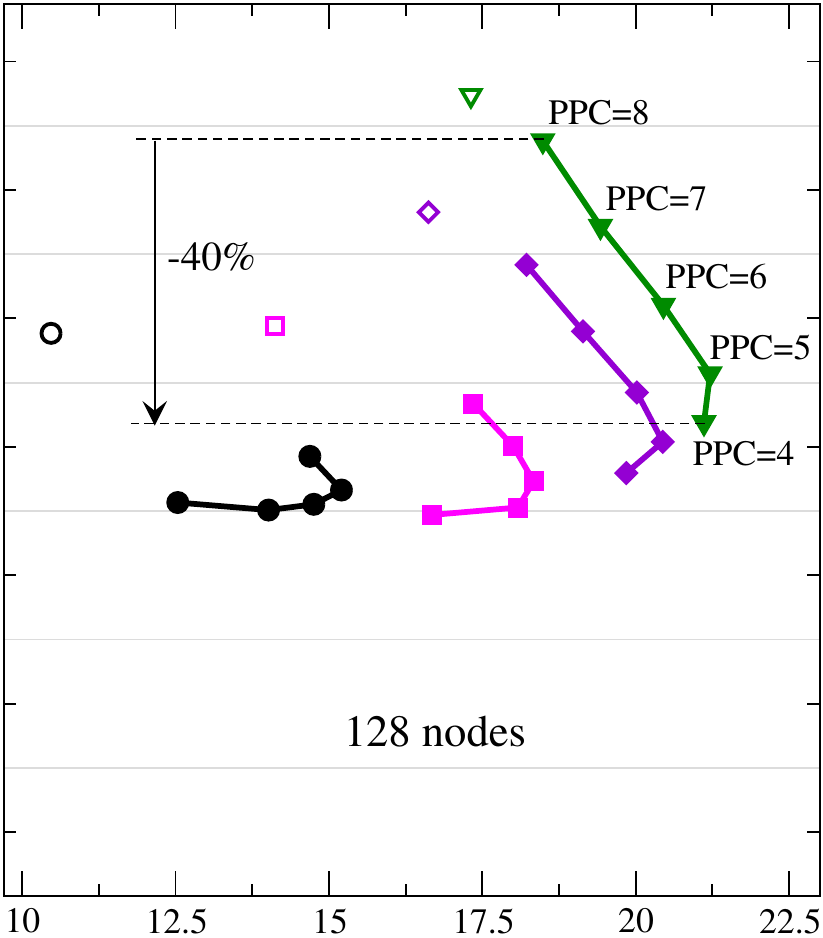}
\centerline{\small Performance [\GFLUPS]}
\caption{\label{fig:ets_perf_packing_scaling}Multi-node energy to
  solution vs.\ performance for the AA pattern AVX LBM implementation
  (large reactor case) for different clock speeds and different node
  counts. The parameter along each curve is the number of processes
  per chip (4~\ldots~8). For comparison, the open symbols show data
  for the scalar implementation on full sockets. Note the overlapping
  scales on the three graphs. The chip-only baseline power of
  $W_0=23\,\W$ was assumed per socket.}
\end{figure}

The question remains whether one can extrapolate the findings about
energy to solution and performance from the chip to the multi-node
level and especially whether single-core optimizations, notably SIMD
vectorization, have a similar
impact. Figure~\ref{fig:ets_perf_packing_scaling} shows aggregated
socket-level energy (as measured via RAPL) vs.\ performance with AVX
for the three node counts (32, 64, and 128) at which parallel
efficiency is between 90 and 60\%. Along each curve, the number of
processes per chip (PPC) is increased from four to eight, and the
highest energy point at the top of each curve is at PPC=8. For
reference we also show the corresponding lowest-energy data points for
the scalar implementation (open symbols).
The overall rise in energy to solution with growing node count
is a trivial consequence of the decreasing parallel efficiency,
because more parallelism means more overhead.
From the data it appears as if a possible optimal operating point
were at 64 nodes, $f=1.2\,\GHZ$, and six cores per chip.
However, performance is about 25\% lower than at $f=2.7\,\GHZ$
with PPC=6 at this node count (see arrow), so the time to solution 
is 33\% larger. This trade-off can only be investigated 
by looking at metrics beyond pure energy to solution, 
such as the energy-delay product or related 
approaches~\cite{bekas:2010,Hsu:2012:TES:2188286.2188309}. Which
metric should be chosen depends on local policies, of course,
and is out of scope for this work.

The most striking difference to the chip-level results is the notable
performance degradation after the saturation point, especially at the
larger node counts (64 and 128). It is caused by the drop in ringshift
bandwidth (as described in the previous section) with growing PPC and
directly leads to a fast rise in energy to solution, much steeper than
would be expected by the power model without communication
component. Hence, it is even more crucial in the highly parallel case
to select the optimal operating point, since each expendable core
costs an over-proportional amount of energy: at 128 nodes and
2.7\,\GHZ, the reduction in energy consumption when going from the
full socket (at PPC=8 with 18.5\,\GFLUPS) to the saturation point (at
PPC=4 with 21.1\,\GFLUPS) is over 40\% (see arrow), but only about 25\% on a
single chip (see the 2.7\,\GHZ\ data in
Fig.~\ref{fig:Z_ets_AAvsScalar}b). 

The strong disadvantage of scalar execution can also be seen
on the highly parallel level (open symbols in 
Fig.~\ref{fig:ets_perf_packing_scaling} show the ``naive'' operating point
of PPC=8 for this case). 
Since more processes are needed to reach saturation -- if this is possible at
all --, the slowdown at larger PPC contributes strongly to the low
performance and high energy consumption.
As a consequence, a well-vectorized LBM code is instrumental 
for optimal energy to solution, particularly in the highly parallel
case when communication plays a noticeable (but not dominant) role.

\begin{figure}[tb]
\centering
\includegraphics*[height=0.53\linewidth]{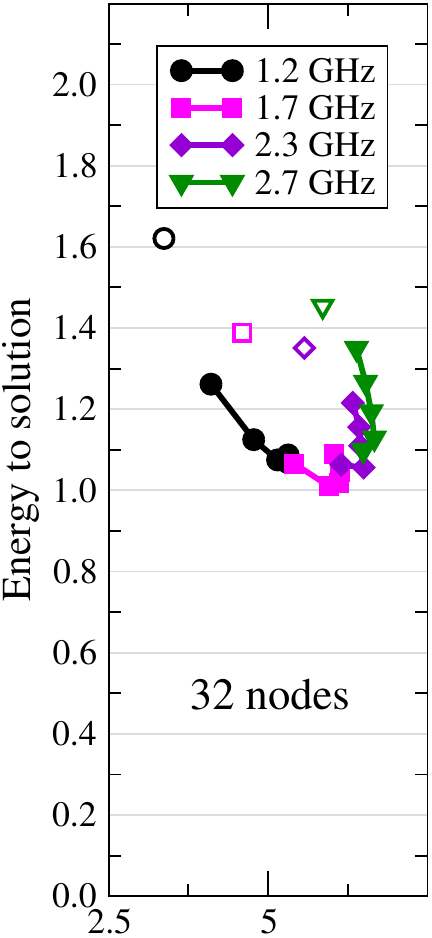}\hfill
\includegraphics*[height=0.53\linewidth]{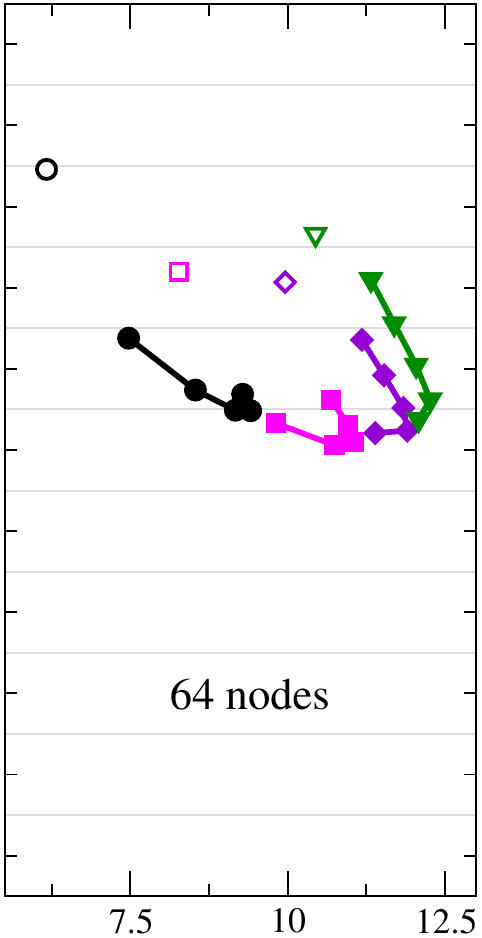}\hfill
\includegraphics*[height=0.53\linewidth]{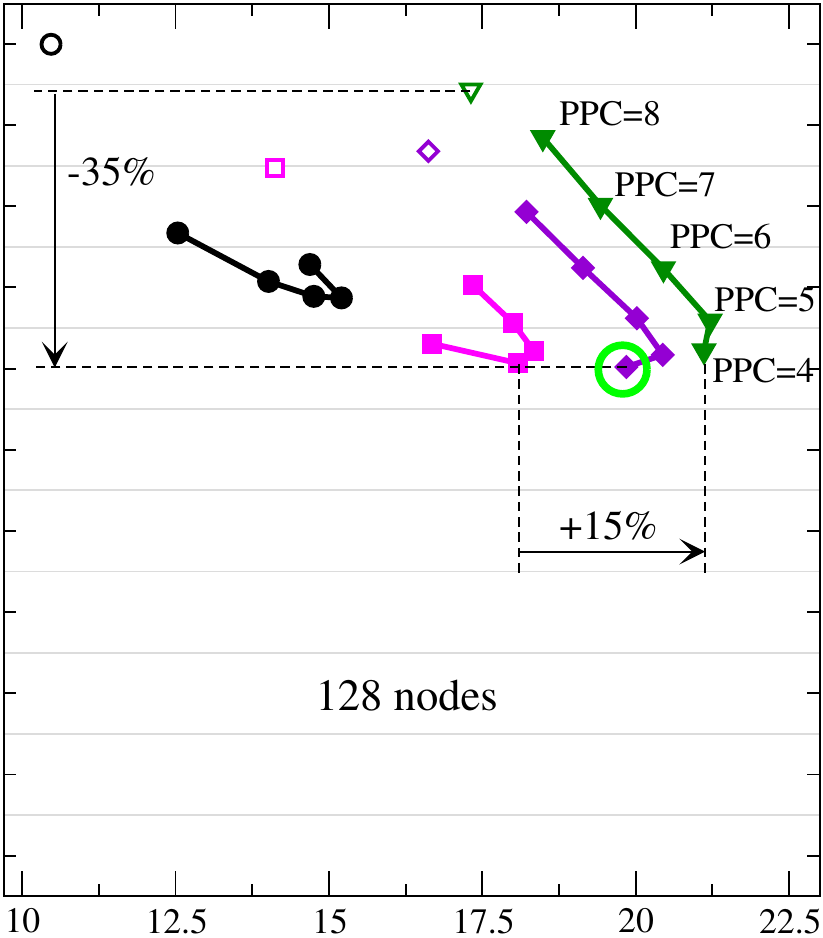}
\centerline{\small Performance [\GFLUPS]}
\caption{\label{fig:ets_perf_packing_scaling_bp}Same data as in
  Fig.~\ref{fig:ets_perf_packing_scaling} but with a realistic
  baseline power of $W_0'=73\,\W$ per socket. 
  The green circle marks a possible
  optimal operating point.}
\end{figure}
Finally, we need to comment on how these findings change if a realistic
baseline power is used. Figure~\ref{fig:ets_perf_packing_scaling_bp}
shows the same data as Fig.~\ref{fig:ets_perf_packing_scaling} but
with the modified per-socket baseline power of $W_0'=73\,\W$ (see
Sect.~\ref{sec:ets_chip}). The results are very similar to the
chip-level discussion: Time to solution becomes more important as the main influencing
factor for energy consumption. All
differences are damped by the larger idle power,
but at 128 nodes there is still about 35\% gain between a naive scalar code
run with PPC=8 on full sockets (open green triangle) 
and the possible optimal operating point, which is marked in
Fig.~\ref{fig:ets_perf_packing_scaling_bp}) at PPC=4 and 2.3\,\GHZ.
In contrast to the case where only the bare chip baseline power 
of $W_0=23\,\W$ is considered, the
lowest frequency setting of 1.2\,\GHZ\ is very unfavorable, not only
in terms of performance but also in terms of energy: the large
performance degradation, the communication bandwidth
breakdown problem, and the large baseline power combined prohibit the use of
very small frequencies, even if energy to solution were the only
relevant metric. On the other hand, energy is practically constant
between 1.7\,\GHZ\ and 2.7\,\GHZ\ when the best PPC value is chosen (PPC=4 at
2.7\,GHZ, PPC=4 (or 5) at 2.3\,\GHZ, and PPC=5 (or 6) at 1.7\,\GHZ),
but performance is boosted by 15\% at 128 nodes (from
18.3\,\GFLUPS\ to 21.1\,\GFLUPS). Hence, when the operating point
is carefully chosen, it is possible to trade
cores for performance at almost constant energy to solution by 
increasing the clock speed and reducing the number of cores
at the same time.

\section{Summary and outlook}\label{sec:conclusion}

We have analyzed the performance and energy to solution properties of
a lattice-Boltzmann flow solver on the chip and highly parallel levels
for an Intel Sandy Bridge EP-based system. Different propagation
patterns (pull-split vs.\ AA), SIMD vectorization schemes (scalar
vs.\ SSE/AVX), and the dependence on the clock speed of the processors
and the number of processes per chip were explained using the
chip-level ECM performance model and a simple power model. In addition
to the measured chip power, the (estimated) ``true'' system-level
power was taken into account in order to arrive at useful predictions
for realistic scenarios. The power model describes well the general
chip-level behavior of the energy to solution metric with
bandwidth-limited loop code: (i) Minimum energy to solution is
achieved at the saturation point. (ii) Lowering the frequency leads to
better scalability across cores and the saturation point is shifted
to larger core counts. (iii) Improved serial performance
leads to fewer cores required to reach saturation.

Since the LBM algorithm and implementation used here is bandwidth-limited
if properly optimized, we 
found a high single-core performance to be the pivotal instrument
for reaching minimal energy without sacrificing too much time to
solution. AVX vectorization and the choice of a good propagation
pattern are important components for accomplishing this goal.  Technical
measures such as clock speed reduction have less impact but still
contribute considerably to the overall energy consumption.

When a ``good'' single-core code has been found and the optimal clock
speed has been set, it is the identification of the performance
saturation point with respect to the number of processes on the chip
level, but more importantly in the highly parallel case, which decides
upon minimal energy to solution. Due to the MPI communication adding a
strongly frequency-dependent component to execution time, using too
many cores per chip must be avoided when communication plays a
non-negligible role. 

The impact of the whole system's baseline power consumption (i.\,e.,
powered on but with idle cores) is twofold: it attenuates the
differences in energy to solution caused by technical measures such as
clock speed adjustments, but it emphasizes the influence of bare
chip-level performance and communication overhead, especially in the
highly parallel case. Hence, a combination of ``good'' single-chip code,
a correct choice of active cores, and an optimal clock speed 
is needed when dealing with a realistic scenario, i.\,e., full-system
power consumption and parallel production runs.
As a consequence, a simple, possibly automatic reduction of the clock speed 
(triggered by the observed strong memory bandwidth utilization) may
only yield a very small fraction of the potential energy savings
compared to a setting in which all parameters are chosen optimally.

These results should be generalizable to any memory-bound algorithm
whose communication overhead becomes significant at strong scaling, if
the behavior of the basic performance-limiting parameters is similar
to the system used in our analysis (\supermuc). Specifically, the
effective communication bandwidth has to depend more strongly on the
clock speed than the memory bandwidth.

This work opens the possibility for future research in multiple
directions. We have not addressed realistic alternatives to the
standard quality metrics of energy and time to solution in detail.
One such cost function could be the energy-delay product or one of 
its variants, but there
may be others. Moreover it will be interesting to compare our findings
for a typical x86-based cluster to a modern low-power system such as
the IBM Blue Gene/Q. Power capping is an additional operational 
constraint of modern systems and can be studied using the same methods
as shown here. Lastly, the ECM model and the multicore power
model need refinements and adjustments to be able to deal with more
complex hardware and software scenarios such as the expected flexible
(per-core) frequency settings on upcoming Intel processors. 
Also it would 
be worthwhile to explore the reasons for the deviation of the ECM model
from measurements near the saturation point, which does not occur
for all types of code \cite{hager:cpe13}.

\section*{Acknowledgments}

We 
gratefully acknowledge the support by LRZ Supercomputing
Center.
This work was partially supported by BMBF under grant No.\ 01IH08003A
(project SKALB).

\bibliographystyle{drgh}
\bibliography{refs}

\end{document}